\documentclass{mn2e}
\usepackage[dvipsnames,usenames]{color}
\usepackage{colortbl}
\usepackage{times}
\usepackage{mathptmx}
\usepackage{epsf}
\usepackage{graphicx}
\usepackage{fixltx2e}
\newcommand{\etal}{{et al}\/.}

\newcommand{\matHI}{\rm H{\hskip 0.02cm\scriptscriptstyle I}}

\newcommand{\erf}{\mathop{\mathrm{erf}}}

\begin{document}
\title[Outflow in 3C\,305]{The nature of the jet-driven outflow in the
radio galaxy 3C\,305}
\author[M.J. Hardcastle \etal]{M.J.\ Hardcastle$^1$\thanks{E-mail:
    m.j.hardcastle@herts.ac.uk}, F.\ Massaro$^{2,3}$, D.E.\ 
  Harris$^2$, S.A.\ Baum$^{4,5}$, S.\ Bianchi$^6$,\newauthor M.\ Chiaberge$^{7,8,9}$, R.\ Morganti$^{10,11}$,
  C.P.\ O'Dea$^{12,2}$, and A. Siemiginowska$^2$\\
$^1$ School of Physics, Astronomy \& Mathematics, University of
  Hertfordshire, College Lane, Hatfield AL10 9AB, UK\\
$^2$ Harvard-Smithsonian Center for Astrophysics, 60 Garden Street,
  Cambridge, MA~02138, USA\\
$^3$ SLAC National Laboratory and Kavli Institute for Particle
Astrophysics and Cosmology, 2575 Sand Hill Road, Menlo Park, CA
94025, USA\\
$^4$ Chester F. Carlson Center for Imaging Science, Rochester
Institute of Technology, Rochester, NY 14623, USA\\
$^5$ Radcliffe Institute for Advanced Study, 10 Garden Street, Cambridge, MA~02138, USA\\
$^6$ Dipartimento di Fisica, Universit\`a degli Studi Roma Tre, via della
Vasca Navale 84, 00146 Roma, Italy\\
$^7$ Space Telescope Science Institute, Baltimore, MD 21218, USA\\
$^8$ Center for Astrophysical Sciences, Johns Hopkins University, 3400 N.
Charles Street Baltimore, MD 21218, USA\\
$^9$INAF - Istituto di Radioastronomia di Bologna, via Gobetti 101 40129
Bologna, Italy\\ 
$^{10}$ ASTRON, P.O. Box 2, 7990 AA Dwingeloo, The Netherlands\\
$^{11}$ Kapteyn Astronomical Institute, University of Groningen, P.O. Box 800,
9700 AV Groningen, The Netherlands\\
$^{12}$ Department of Physics, Rochester Institute of Technology, 54 Lomb Memorial Drive, Rochester, NY 14623, USA\\
}
\maketitle
\begin{abstract}
We present {\it Chandra} X-ray and VLA radio observations of the radio
galaxy 3C\,305. The X-ray observations reveal the details of the
previously known extended X-ray halo around the radio galaxy. We
show using X-ray spectroscopy that the X-ray emission is
consistent with being shock-heated material and can be modelled
with standard collisional-ionization models, rather than being
photoionized by the active nucleus. On this basis, we can make a
self-consistent model in which the X-ray-emitting plasma is
responsible for the depolarization of some regions of the radio
emission from the jets and hotspots, and to place lower and upper
limits on the magnetic field strength in the depolarizing medium. On
the assumption that the X-ray-emitting material, together with the
previously-known extended emission-line region and the outflow in
neutral hydrogen, are all being driven out of the centre of the galaxy
by an interaction with the jets, we derive a detailed energy budget
for the radio galaxy, showing that the X-ray-emitting gas dominates
the other phases in terms of its energy content. The power supplied by
the jets must be $\sim 10^{43}$ erg s$^{-1}$.
\end{abstract}

\section{Introduction}
\label{intro}

In powerful radio galaxies, extended optical emission-line regions
(EELR) are often found to be aligned with the axis defined by the
extended radio emission (e.g., McCarthy \etal\ 1987; McCarthy 1993;
McCarthy \etal\ 1996). A long-standing question (e.g., Baum \& Heckman
1989; Tadhunter \etal\ 1998) is whether these EELR are ionized by
photons from the nucleus or by shocks driven by the jets; in the most
powerful radio sources, normally found at high redshift, there is
often evidence that the optical emission-line material is either ionized or
at least strongly affected by jet-driven shocks, which requires a
direct interaction between the jets and the warm/cold ($T < 10^4$ K) phase
of the ISM of the host galaxy (e.g., S\'olorzano-I\~narrea
\etal\ 2002; Nesvadba \etal\ 2008).

Where there is evidence for interaction between the jets and the
optical emission-line properties, the optical lines can be used to
probe the {\it dynamics} of the material interacting with the radio
source, since accurate radial velocities can be measured. It then
becomes clear in many cases that the radio source drives a large-scale
outflow, at least in the warm-gas phase, with typical speeds of
hundreds of km s$^{-1}$ (e.g., O'Dea \etal\ 2002). The mass of the
emission-line material itself is small, but its filling factor is
necessarily very low, so that it is probably being driven out together
with a much larger mass of material at some other temperature or
ionization state. At least two components of this additional material
must be present. On the one hand, it has recently been shown (e.g.,
Morganti, Tadhunter \& Oosterloo 2005b) that at low redshift these
sources often drive large-scale outflows of neutral hydrogen (i.e.,
colder gas) with very similar velocities. On the other,
observations of radio depolarization associated with the extended
emission-line regions (Heckman \etal\ 1982, 1984; Hardcastle
\etal\ 1997) require the presence of a massive, ionized (i.e.,
warm--hot), magnetized medium in which the emission-line clouds are
embedded (Hardcastle 2003). The precise mechanism of the interaction
between the radio jets and these other phases of the ISM is not clear,
but it makes radio-galaxy driven extended emission-line regions into
objects of great interest for models of galaxy formation and
evolution, since they provide a {\it kinetically}, rather than
radiatively, driven method by which the central AGN can expel
cold/warm gas from the center of the host galaxy (e.g., Tadhunter
2007).

The most recent clues to the nature of these interactions have come
from X-ray observations with {\it Chandra}. In Hardcastle
\etal\ (2010; hereafter H10) we presented observations of the
intermediate-redshift, powerful radio galaxy 3C\,171 ($z=0.2384$),
which has long been known to be associated with an emission-line
outflow and with strong radio depolarization (e.g., Clark \etal\ 1998;
Hardcastle \etal\ 2003). The {\it Chandra} observations revealed an
extended X-ray region that was close to co-spatial with the
emission-line material as seen in [O{\sc iii}], [O{\sc ii}] and H$\alpha$,
and also very well aligned with the region of radio depolarization.
Morphologically this allowed us to rule out non-thermal (e.g., X-ray
synchrotron) models for the X-rays. In interpreting the physics of
X-rays associated with emission-line regions it is important to try to
distinguish between photoionized and collisionally ionized
X-ray-emitting material, since it is known (e.g., Bianchi \etal\ 2006)
that in local Seyfert galaxies photoionization dominates. This was
impossible in 3C\,171 from direct observations of the X-ray spectrum
(although photoionization was disfavored by morphological and
energetic arguments) but we were able to estimate a temperature and
density in the extended structures on the assumption of a
collisional-ionization model, and then to show that this was consistent
with the properties of the line-emitting gas and with the constraints
from radio depolarization. A photoionization model would not be
consistent with these observations. We were therefore able to estimate
the mass of the X-ray-emitting material and to show that, on the
assumption that this material is shocked gas flowing out along with
the emission-line clouds, it completely dominates the energetics of
the outflow; the total power in the outflow over the source lifetime
was $\sim 3 \times 10^{44}$ erg s$^{-1}$, comparable to the bolometric
radiative power of the AGN. We argued that 3C\,171 is an analogue,
presumably driven by a recent gas-rich merger, of the somewhat more
powerful and much more common systems with extended X-ray emission and
optical emission lines at high redshift (e.g., PKS 1138$-$262 at
$z=2.2$, Carilli \etal\ 2002) but with the advantage of being at low
redshift and thus capable of being studied at high spatial resolution.

In this paper we discuss similarly motivated {\it Chandra} and radio
observations of the nearby radio galaxy 3C\,305. 3C\,305 is a
low-redshift radio galaxy which has been extensively studied because
of its peculiar radio structure (Heckman \etal\ 1982, hereafter H82):
its bright jets and hotspots are embedded in a pair of lobes which
are much longer perpendicular to the jet axis than the core-hotspot distance.
Its 178-MHz radio luminosity is $5.5 \times 10^{24}$ W Hz$^{-1}$
sr$^{-1}$, which puts it right on the FRI/FRII luminosity boundary of
Fanaroff \& Riley (1974). On the strict morphological definition, its
distorted radio structure (Heckman \etal\ 1982) makes it an FRI (that
is, the brightness in a low-resolution map would be peaked closer to
the centre than the edges of the structure) and it is classed as such
in catalogues such as that of Laing, Riley \& Longair (1983). However,
high-resolution radio imaging (e.g. Jackson \etal\ 2003, and this paper)
clearly shows it to have jets and hotspots that are more
characteristic of FRIIs.

3C\,305 is similar to 3C\,171 in many ways: it is a powerful radio
galaxy of unusual morphology that is well known to be associated with
an extended emission-line region on scales of the jets/hotspots, and
to have strong co-spatial radio depolarization (H82); there is strong
evidence in the case of 3C\,305, e.g., from its morphology and from
the detection of [Fe{\sc ii}] emission peaked around the jet
termination point, that the optical emission-line region is
collisionally ionized (e.g., Jackson \etal\ 2003); and, crucially, it
has a known association between the emission-line gas and X-ray
emission on scales that are well resolved by {\it Chandra} (Massaro
\etal\ 2009, hereafter M09; Fig.\ 1), discovered, like 3C\,171's, as a
result of the `snapshot survey' of 3CR radio galaxies (Massaro
\etal\ 2010). However, it differs in four key respects: 1) it is
significantly lower in redshift ($z=0.0416$), and physically smaller,
so better matched to the scales of the host galaxy ISM; 2) its bright,
relatively isolated host galaxy, IC 1065, is relatively well-studied
and is known to show strong morphological and dynamical peculiarities
consistent with being a recent merger involving at least one gas-rich
spiral (Heckman \etal\ 1985); 3) it exhibits one of the best-studied
examples of an association between the radio jets and an H{\sc i}
outflow (Morganti \etal\ 2005a); and 4) it is significantly (a factor
$\sim 4$) brighter in extended X-rays than 3C\,171. 3C\,305 could be
considered as a prototypical example of a young, galaxy-scale powerful
radio source driven by galaxy-galaxy interactions, as expected in
models of galaxy formation in the early universe, and, as emphasised
by Morganti \etal\ (2005b), the presence of both warm and cold phases
of the gas in the fast outflow already represents a challenge for
models describing the interaction between radio jets and their
environment.

In the present paper we present new {\it Chandra} and radio
observations (Section 2) which give us the most sensitive X-ray image,
and the most sensitive and highest-resolution radio images yet made of
3C\,305. We show that the depolarized regions of radio emission become
polarized again at high frequency, allowing the techniques developed
by H10 to be applied to this source (Section 3). The new {\it Chandra}
images confirm that the extended X-ray emission is morphologically
very similar to the previously known emission-line nebula, but also
show some important differences (Section 4); there is no evidence for
large-scale X-ray emission from a group or cluster environment, and
the X-ray spectrum of the extended emission is consistent with being
thermal emission from the ISM of the host galaxy shocked by the jets.
In Section 5 we combine the available data to show that it is
plausible that the X-ray-emitting gas is the depolarizing medium, and
use the picture of the source implied by these measurements to
estimate its energy budget. Our conclusions are presented in Section
6.

Throughout this paper we use a cosmology with $H_0 = 70$ km
s$^{-1}$ Mpc$^{-1}$, $\Omega_{\rm m} = 0.3$ and $\Omega_\Lambda =
0.7$. This gives a luminosity distance to 3C\,305 of
184 Mpc and an angular scale of 0.821 kpc arcsec$^{-1}$.

\section{Observations and data reduction}

\subsection{Radio observations}

Deep radio observations of 3C\,305 have previously been made with the
VLA (H82) and MERLIN (Jackson \etal\ 2003). Our motivation in making
new observations was to obtain a high-resolution, high-frequency view
of the polarized and total intensity in order to constrain the
properties of the depolarizing medium: the essentially complete
depolarization observed across broad regions of the source by H82
implies a significant Faraday-active medium but only gives limits on
its properties. We therefore observed 3C\,305 with the VLA (proposal
ID AH982) at frequencies around 8.6 GHz (X-band) and 22.3 GHz (K-band),
the highest pair of frequencies that would give adequate sensitivity
at the time of the proposal, using several VLA configurations;
observational details are given in Table \ref{vla-obs}. All
observations used two observing frequencies each with 50 MHz bandwidth
(the two frequencies used are given in Table \ref{vla-obs}). For both
observations, we used primary referenced pointing to obtain the best
possible pointing accuracy of the array -- this is particularly
important at K-band where the primary beam of the VLA is small (FWHP 2
arcmin). For the K-band observations we used fast-switching mode to
nod between the target and a nearby calibrator, 1436+636, 1.4$^\circ$
from the target, on a timescale of 100 s (70 s on source, 30 s on
calibrator). This allows accurate tracking of the rapidly varying
atmospheric phases at K-band. The short baselines provided by the more
compact configurations of the VLA (the shortest baseline at both bands
is $\sim 6$ k$\lambda$, corresponding to an angular scale of 30
arcsec) means that the whole source (LAS = 13 arcsec) is adequately
sampled.

All observations were reduced in AIPS in the standard manner.
Polarization calibration was carried out using 3C\,286 as a reference
source. After initial imaging, we carried out several iterations of
phase self-calibration on all but the 22.3-GHz A-configuration data
(where the signal-to-noise was too low to give a significant
improvement over the already good phases provided by fast switching).
After phase self-calibration and cross-calibration to ensure that
phases were aligned, the $uv$ datasets were concatenated for imaging.

\begin{table}
\caption{Observations of 3C\,305 with the VLA}
\label{vla-obs}
\begin{tabular}{llrr}
\hline
Date&Config&Frequencies (GHz)&Time on source (h)\\
\hline
2008-Oct-26&A&22.135, 22.535&2.6\\
&&8.435, 8.735&2.0\\[2pt]
2009-Apr-14&B&22.135, 22.535&2.6\\
&&8.435, 8.735&2.0\\[2pt]
2009-Jul-03&C&22.135, 22.535&2.8\\
\hline
\end{tabular}
\end{table}

\begin{figure*}
\epsfxsize 14cm
\epsfbox{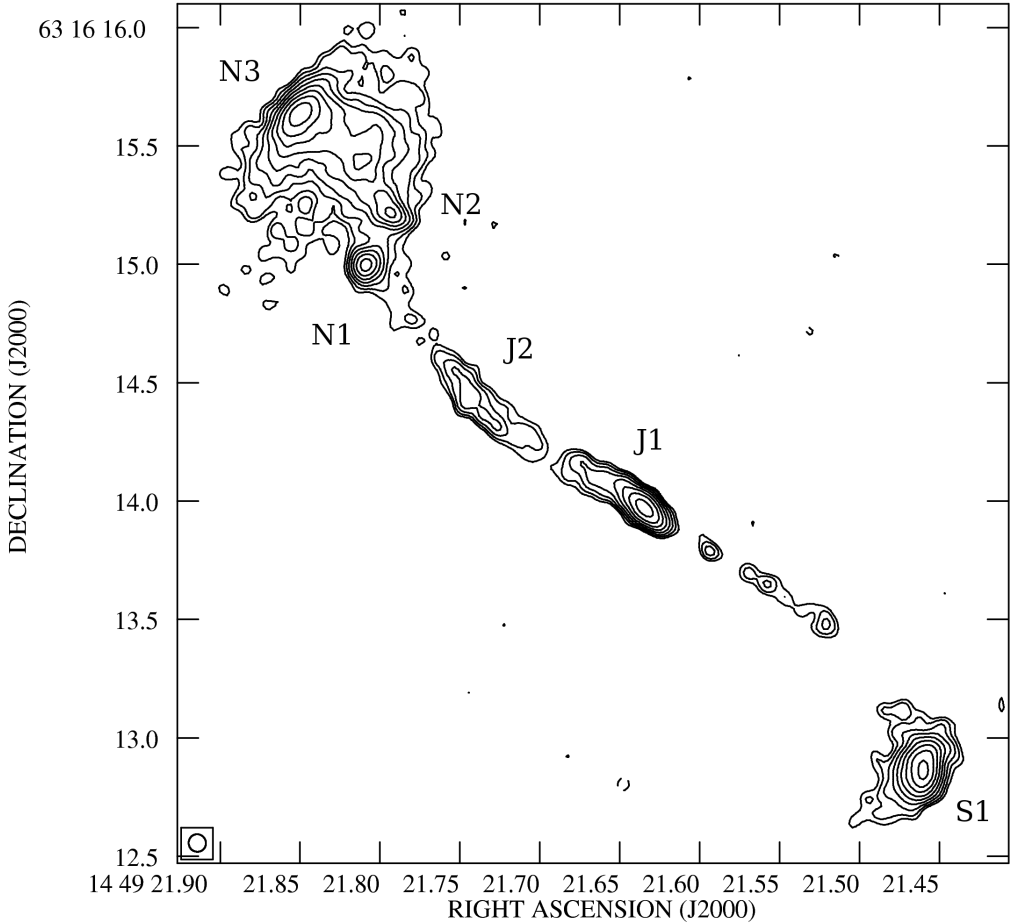}
\caption{Full-resolution ($89 \times 84$ mas) total-intensity image of
  3C\,305 at K-band (22.3 GHz). Contours start at the $4\sigma$ level as given in
  the text and increase by a factor $\sqrt{2}$ at each step. Features
  discussed in the text are labelled. The distance from hotspot
    to hotspot (peak of N3 to peak of S1) is 3.9 arcsec (3.2 kpc).}
\label{k-fullres}
\end{figure*}

\begin{figure*}
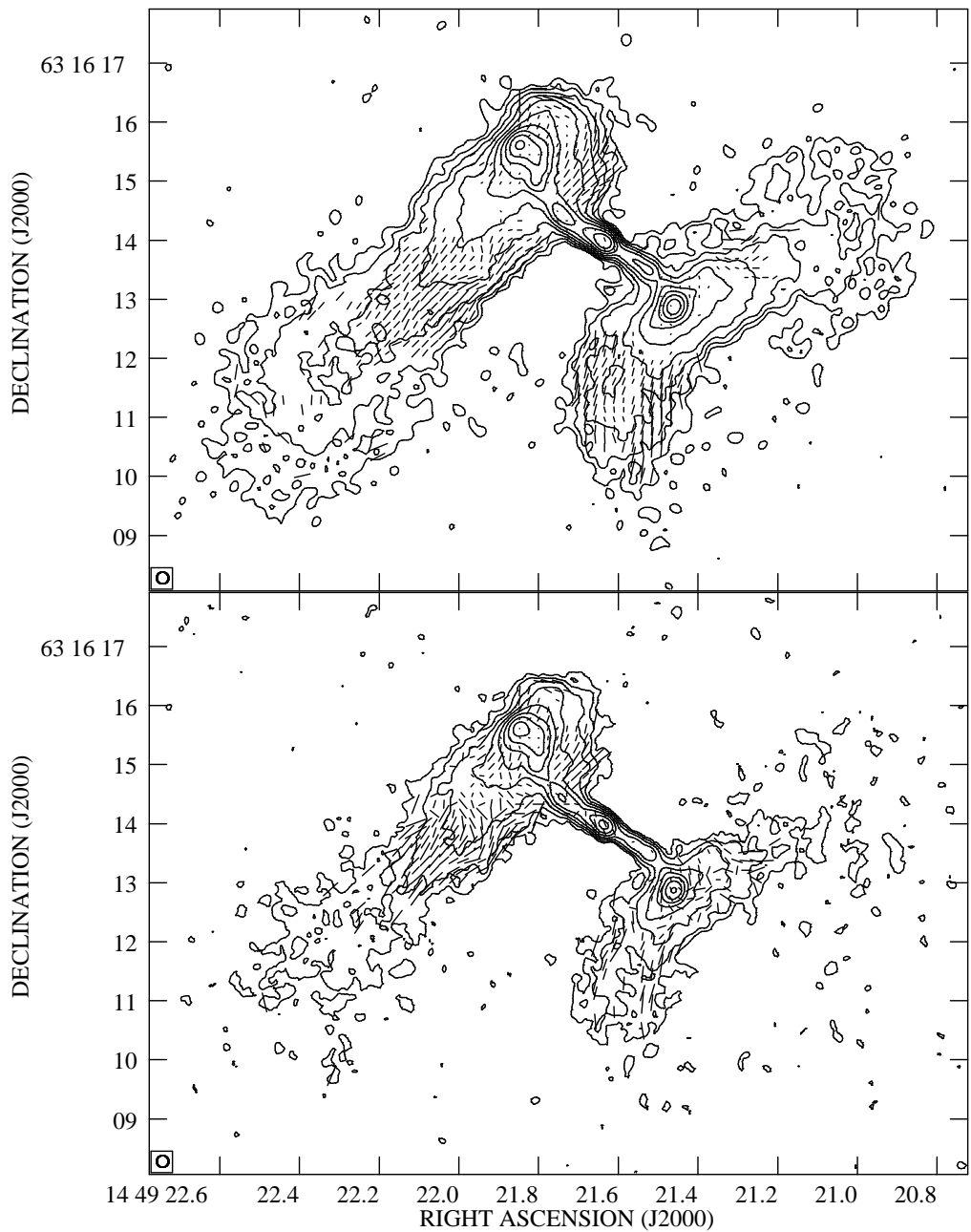

\epsfxsize 14cm
\epsfbox{3C305.XMATCH-fixed.PS}
\vskip -37.5pt
\epsfxsize 14cm
\epsfbox{3C305.KMATCH.PS}
\caption{Matched-resolution VLA images of 3C\,305 at (top panel)
  X-band (8.6 GHz) and (bottom
panel) K-band (22.3 GHz). Contours start at the $3\sigma$ level,
as given in the text, and increase by a factor 2 at each step. Polarization vectors
are plotted at $90^\circ$ to the E-vector direction, and their length
shows relative fractional polarization. Vectors are only plotted where
the signal in both polarization and total intensity exceeds the
$3\sigma$ level.}
\label{vla-maps}
\end{figure*}

Imaging was carried out in AIPS using suitable tapering and weights to
obtain appropriate resolutions. Our final imaging products were a
full-resolution K-band image (resolution $89 \times 84$ mas: rms noise
32 $\mu$Jy beam$^{-1}$ in all three Stokes parameters), shown in
Fig.\ \ref{k-fullres}, and
matched-resolution, matched-shortest-baseline X-
and K-band images (resolution $0.2 \times 0.2$ arcsec: rms noise 17
$\mu$Jy beam$^{-1}$ at X-band and 23 $\mu$Jy beam$^{-1}$ at K-band),
which are shown in Fig. \ref{vla-maps}.

We also obtained the L-band (1.4 GHz) MERLIN observations of Jackson
\etal\ (2003) from the MERLIN archive, using the standard archive
processing, and re-processed a short archival A-configuration L-band
VLA observation, with the aim of searching for any polarization at
L-band. We do not present any images from these data (which are the
basis of the 3CRR Atlas\footnote{http://www.jb.man.ac.uk/atlas/}
images used by M09) but comment on their implications for
depolarization in Section \ref{depol}.

\subsection{{\it Chandra} observations}
\label{astrometry}

3C\,305 was observed with {\it Chandra} for 8 ks in 2008 (M09); our
two new observations were taken in early 2011. All three observations used
the ACIS-S instrument in VFAINT mode with the radio galaxy lying at
the aim point on the back-illuminated S3 chip. Observational details
are given in Table \ref{chandra-obs}. There were no periods of
significantly high background, and so the livetimes quoted are
unfiltered. All three observations were reprocessed using CIAO 4.3 and
CALDB 4.4.5 in the standard manner to apply the latest calibration and
to apply `VFAINT cleaning' to reduce the cosmic-ray background.

\begin{table}
\caption{{\it Chandra} observations of 3C\,305}
\label{chandra-obs}
\begin{tabular}{lrr}
\hline
Date&Obsid&Livetime\\
\hline
2008 Apr 07&9330&8218\\
2011 Jan 03&12797&28661\\
2011 Jan 06&13211&28661\\
\hline
\end{tabular}
\end{table}

For imaging, we used the CIAO {\it merge\_all} script to produce a
merged events file in the energy range 0.5-5.0 keV, with a total
  exposure time of 65.5 ks. We generate images from this events
  file by binning: as the best estimates of the arrival positions of
  individual photons have sub-pixel accuracies, due to the multi-pixel
  nature of event reconstruction and the dithering of the spacecraft,
  we generally use sub-pixel binning and smoothing to produce the best
  images.

In our spectral analysis, we extracted spectra in the standard manner
using the CIAO {\it specextract} script. Spectra from each individual
observation, and their associated response files, were merged using
the {\it combine\_spectra} contributed tool before fitting, as this
makes the best use of the data when, as in this case, the dataset is
broken up into several small pieces. The merged spectra were then
binned so as to have a minimum of 20 counts per bin after background
subtraction, and spectral fitting was carried out in XSPEC 12. For all
our spectral fitting we assumed a Galactic column density of hydrogen,
$N_{\rm H} = 1.69 \times 10^{20}$ cm$^{-2}$, as given by the online
COLDEN\footnote{http://cxc.harvard.edu/toolkit/colden.jsp} tool. Fits
were carried out in the energy range 0.4-7.0 keV and errors quoted are
$1\sigma$ for one interesting parameter.

Accurate astrometric alignment of the X-ray and radio data is
difficult. Unusually for a radio galaxy of this luminosity, 3C\,305
shows no central unresolved radio component; the brightest central
feature of the radio maps is resolved, strongly polarized and
steep-spectrum (as noted by Jackson \etal\ 2003, and see
Fig.\ \ref{vla-maps}) and so it is not appropriate to align the maps
on this feature. There is also no clear unresolved central point
source in the {\it Chandra} data, suggesting that any AGN- or
jet-related X-ray emission is weak (as previously discussed by Evans
\etal\ 2009). If we consider only the hard X-ray emission from the
source ($>4$ keV), which might provide a clue to the location of a
heavily absorbed central AGN, we find that there is a significant
detection of the central parts of the X-ray emission, whose centroid
with the default astrometry is located just at the base of the
brightest radio emission, though as there are only 17 counts in a
1-arcsec source circle in the combined image above 4 keV the errors in
this position are necessarily large. With this caveat, given that the
centroid of these counts lies in a very plausible position for the
hidden AGN, we use the default astrometry to relate the X-ray and
radio images. This is slightly different from the approach taken by
M09, because we have enough hard counts to at least tentatively locate
the AGN by this method and because we do not attempt to align on the
base of the jet.

\begin{figure*}
\epsfxsize 17cm
\epsfbox{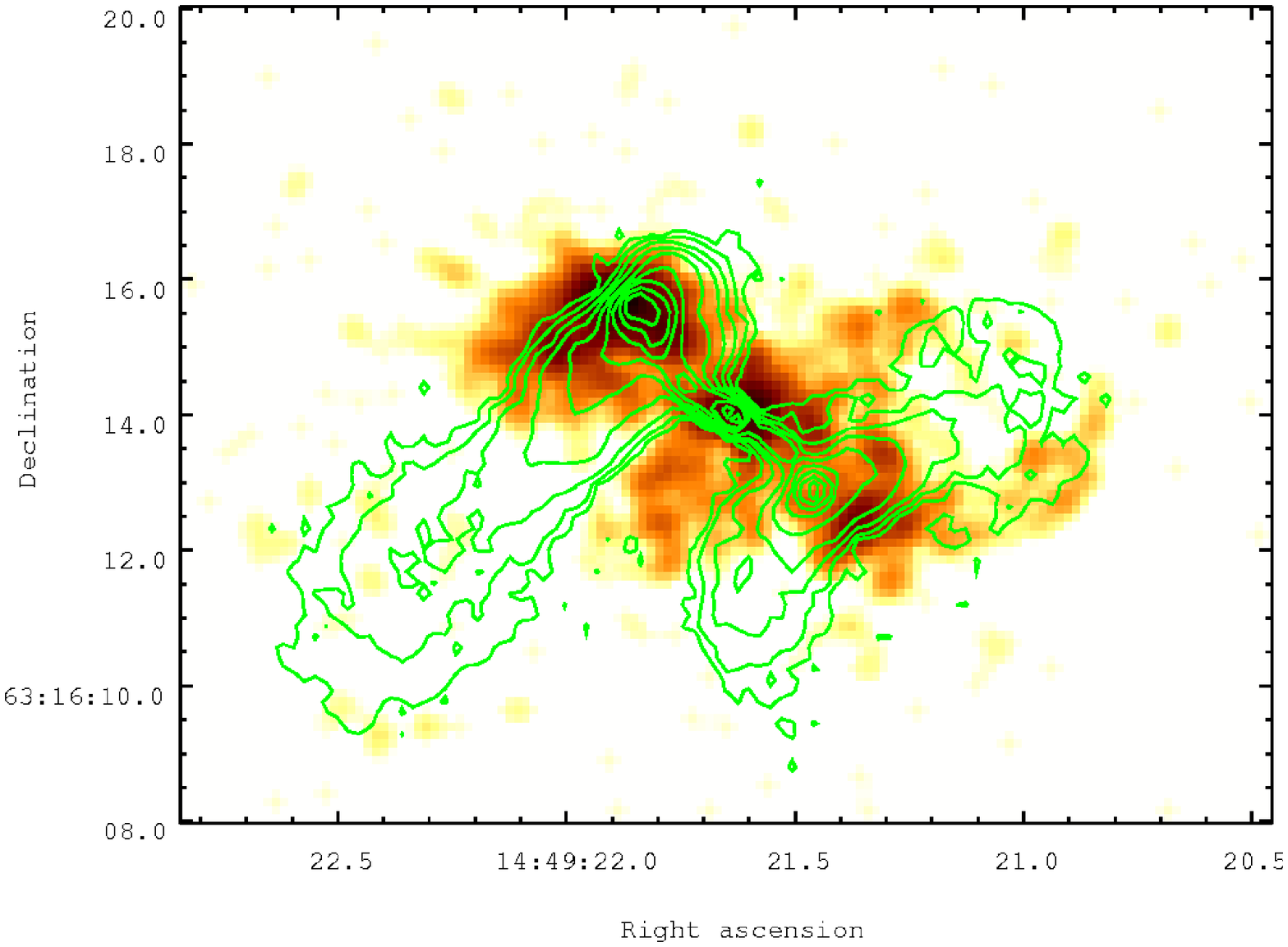}
\caption{Radio and X-ray image of 3C\,305. Contours are from the 8-GHz
image of Fig.\ \ref{vla-maps}. The colours show the {\it Chandra} data
in the 0.5-5.0 keV band, binned in pixels of 0.123 arcsec on a side and
smoothed with a Gaussian of FWHM 3 pixels; the smoothed {\it Chandra}
image has an effective
resolution of $\sim 0.6$ arcsec.}
\label{chandra-overlay}
\end{figure*}

An overlay of the radio and X-ray images, showing the key features of
the X-ray data, is shown in Fig.\ \ref{chandra-overlay}.

\subsection{Other data}

\begin{figure*}
\epsfxsize 17cm
\epsfbox{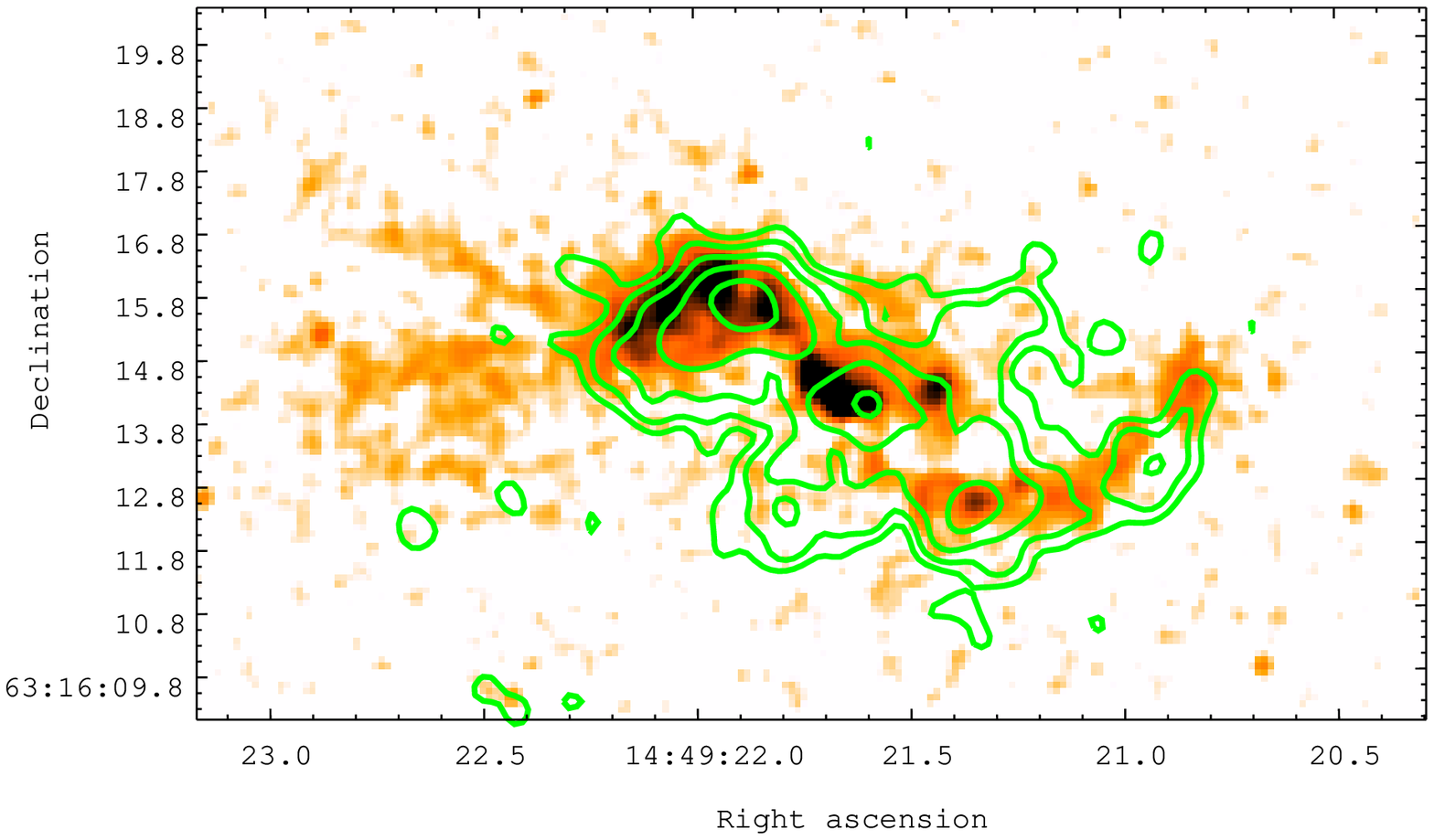}
\caption{[O{\sc iii}] and X-ray image of 3C\,305. Contours show the 0.5-5.0
  keV X-ray emission, binned and smoothed as in
  Fig.\ \ref{chandra-overlay}: the lowest contour is the $3\sigma$
  surface-brightness level, calculated using the method of Hardcastle
  (2000), relative to the off-source background, and each successive
  contour is a factor 2 higher in surface brightness. The colours show the
  [O{\sc iii}] 500.7-nm image of Privon \etal\ (2008), smoothed with a
  Gaussian of FWHM 2 pixels to improve surface-brightness sensitivity.}
\label{oiii-overlay}
\end{figure*}

We also make use of the {\it Hubble Space Telescope} ({\it HST})
continuum-subtracted image of the 500.7 nm [O{\sc iii}] line, originally
presented by Privon \etal\ (2008) and also used by M09; this
  image's astrometry was corrected by M09 to align the peak of the
  [O{\sc iii}] emission with the base of the jet in the radio map, and
  we adopt this alignment in the absence of any more accurate location
of the AGN in either map. The
relationship between the [O{\sc iii}] and X-ray emission is shown in
Fig.\ \ref{oiii-overlay}. It can be seen that there is not a
one-to-one correspondence between the X-ray and [O{\sc iii}] emission,
though there are many features in common. The X-ray emission appears
more similar to the image of H$\alpha$+[NII] presented by H82 (their
fig.\ 4).

\section{Radio results}

\subsection{Total intensity}

Our lower-resolution images (Fig.\ \ref{vla-maps}) show the structure
of the source well and give a very similar picture to that observed in
earlier VLA and MERLIN imaging (H82; Jackson \etal\ 2003). Like many
low-redshift radio galaxies, 3C\,305 has obvious bright jets and
hotspots, but, unlike most, it has low-surface brightness,
steep-spectrum regions that extend perpendicular to the source on
either side of the jet. These regions (which we will refer to as the
lobes, in spite of their unusual morphology) are strongly polarized
with the magnetic field vector pointing down the surface brightness
gradient (i.e. perpendicular to the jet).

The jet is well resolved at the full K-band resolution
(Fig.\ \ref{k-fullres}) and clearly shows a slight S-shape in both jet
and counterjet, with edge-brightening at the outer edges of curves
(J1, J2). A bright knot at the end of the jet, N1, is the most compact
feature in the hotspot region, and would be described as the primary
hotspot by the criteria of Hardcastle \etal\ (1997), meaning that the
brightest part of the NE lobe (N2, N3, and the bright emission around
them) is likely to be a secondary hotspot complex characterized by
repeated redirection of the jet flow. This is consistent with the fact
that N1, N2 and N3 are all elongated perpendicular to the presumed
local jet direction. By contrast, there is only a single compact
feature, S1, in the SW lobe, again clearly elongated perpendicular to
the jet direction.

\subsection{Polarization}
\label{depol}

\begin{figure*}
\epsfxsize 17cm
\epsfbox{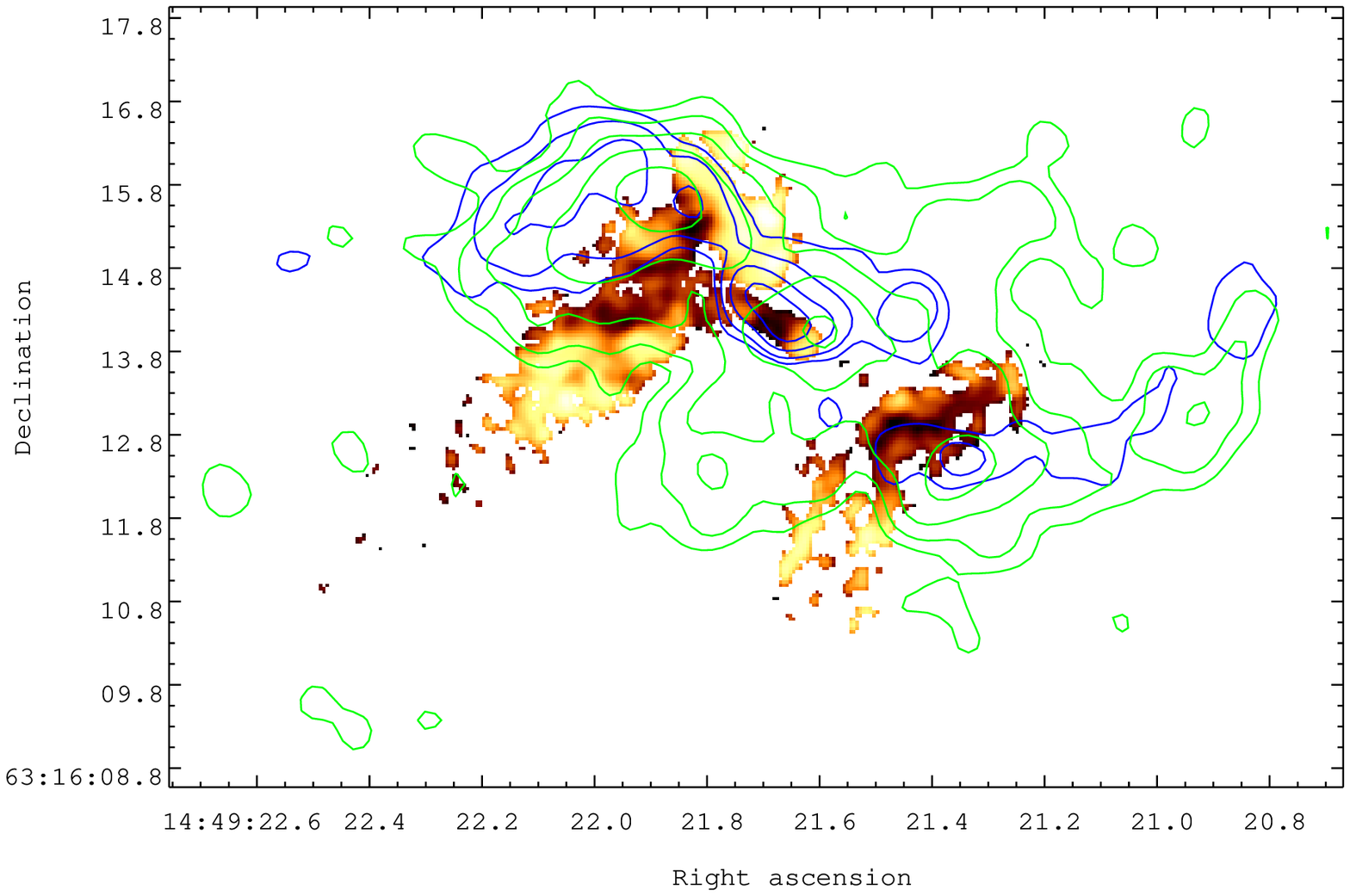}
\caption{A map of the depolarization measure $DP_8^{22}$ between 22.3
  and 8.6 GHz, as described in the text. Overlaid are contours of the
  X-ray emission (green, as in Fig.\ \ref{oiii-overlay}) and the
  smoothed [O{\sc iii}] emission from Fig.\ \ref{oiii-overlay} (blue). Black
  regions have $DP<0.1$ (highly depolarized), red-orange have $0.1 <
  DP < 0.5$ (moderately depolarized) and
  yellow-white regions show little or no depolarization ($0.5 < DP \la
  1$).}
\label{dp-overlay}
\end{figure*}

As first noted by H82 there are regions of strong
depolarization close to the jet axis in the lobes and hotspots of
3C\,305. Our 8-GHz image, though made with very much better
sensitivity and resolution, shows very similar polarization properties
to the 5-GHz image presented by H82, including regions
where little or no polarization is detected (Fig.\ \ref{vla-maps}). At
22.3 GHz, though, almost all of the bright regions of the source are
strongly polarized, as we would expect for the hotspot regions of a
radio galaxy. This frequency dependence, in maps of matched spatial
resolution, rules out simple beam depolarization, where the
depolarization arises because we are averaging over unresolved complex
polarization structure.

Because we see repolarization at high frequencies we can characterize
the depolarization quantitatively. We calculate the quantity
$DP_8^{22}$, defined as the ratio of the fractional polarization at
8.6 GHz to that at 22.3 GHz, at each pixel in the map where the total
intensity at both 8.6 and 22.3 GHz exceeds the $4\sigma$ level and
where the polarized intensity at 22.3 GHz also exceeds the $4\sigma$
level. Where there is no detected polarization at 8.6 GHz in a pixel
that satisfies these criteria, we replace the 8.6-GHz polarization
with its $4\sigma$ upper limit so that the depolarization can be
easily visualised. The resulting map of $DP_8^{22}$, with overlaid
[O{\sc iii}] and X-ray contours, is shown in Fig.\ \ref{dp-overlay}.
This image has several interesting features. First, we note that the
depolarization is at very similar levels in the NE and SW lobes: since
there are several arguments (Jackson \etal\ 2003) that the source is
at a non-negligible angle to the line of sight, this gives an
indication that any depolarizing medium must be closely associated
with the direction of the radio jet rather than being an unrelated
foreground feature in the galaxy, since otherwise we would expect to
see a jet/counterjet side asymmetry along the lines of the well-known
Laing-Garrington effect (Laing 1988; Garrington \etal\ 1998).
Secondly, we see strong, resolved variations in $DP$ even in the
depolarized regions (especially when we bear in mind that in some
cases we are only measuring an upper limit on $DP$): although some of
this may be attributable to differences in the geometry of the
depolarizing medium or to different intrinsic polarizations at
different frequencies, it seems most likely that the depolarizing
medium must be clumpy on scales of $\sim 0.5$ arcsec (0.4 kpc).
Thirdly, we see that in the NE lobe the edges of the depolarization
region are quite clearly delineated -- the lobe to the NW and SE of
the jet has $DP \sim 1$ -- and that these edges are not parallel to
the jet direction. The depolarization silhouette within the lobe is
only 0.8 arcsec (0.7 kpc) wide at the innermost point at which this
can be measured, compared to 2.0 arcsec (1.6 kpc) at its outermost
point. We cannot say definitively whether this is so in the SW lobe
because the NW edge of this feature is not very well detected at 22.3
GHz, presumably due to its steep radio spectrum, but there are hints
that the depolarization is lower on the NW edge. Finally, we note that
the depolarization seems to be considerably better associated
  with the X-ray emission than with the emission-line gas as traced
by the [O{\sc iii}] image -- in particular, all depolarized parts
  of the source are coincident with bright X-ray emission, while
there is a region in the NE lobe where there is strong depolarization
but little or no [O{\sc iii}] emission. All of these features, taken
together, are qualitatively very consistent with the idea that the
depolarization is due to a Faraday screen, closely associated with but
external to the jets and lobes, and which either is the X-ray-emitting
gas itself or is distributed in a very similar way. We will test this
idea more quantitatively in the following sections.

Finally, we note that no polarized emission is seen from the L-band
(1.4 GHz) MERLIN observations, and almost none from the
A-configuration L-band VLA data -- there is a weak detection of
polarization from the NW edge of the NE lobe, corresponding to the
region in Fig.\ \ref{dp-overlay} where $DP_8^{22} \approx 1$. As the
outer parts of the lobes are strongly and uniformly polarized at 8
GHz, this suggests that there is some Faraday-active medium on larger
scales than the observed X-ray emission.

\section{X-ray results}

\subsection{Imaging}
\label{environment}

As seen in Fig.\ \ref{chandra-overlay}, and as previously noted by
M09, the X-ray emission is morphologically different from the radio
emission and extended on a slightly larger scale. We detect the bright
features associated with the core or inner jet and peaking just
outside the two radio-bright hotspots, but our much deeper X-ray
images (we see $1080\pm30$ 0.5-5.0 keV net counts in a 15-arcsec
source circle) reveal lower-surface-brightness diffuse emission lying
between the two lobes (the `wings') and just outside the SW lobe (the
`ridge'). There is a striking similarity to the [O{\sc iii}] emission
in many places, for example in the ridge (Fig.\ \ref{oiii-overlay}),
but also some differences: in particular, the `wings' between the two
lobes have no [O{\sc iii}] counterpart in the {\it HST} images,
although, as noted above, there may be some corresponding H$\alpha$ +
[N{\sc ii}] emission in the ground-based images of H82. Similar X-ray
and H$\alpha$ extension perpendicular to the jets, though on a larger
physical scale, was seen in our observations of 3C\,171.

We find no evidence for the larger-scale X-ray emission (beyond 15
arcsec, or 12 kpc, from the core) that would be expected if 3C\,305
lay in a cluster or group of galaxies. We estimate a total of $90 \pm
70$ 0.5-5.0 keV counts in an annulus between 15 arcsec and the edge of
the S3 chip, at 1.5 arcmin (73 kpc); this corresponds to a $3\sigma$ upper
limit on bolometric luminosity around $10^{41}$ erg s$^{-1}$ (for $kT
= 1$ keV, abundance 0.3 solar), which is well below the total
luminosities of typical groups (e.g. Osmond \& Ponman 2004) and, as
will be seen below, considerably lower than the luminosity of the
extended X-ray emission within 15 arcsec.  As IC 1065 is a low-redshift
galaxy located in the part of the sky covered by the Sloan Digital Sky
Survey (SDSS), we searched, using the NASA Extragalactic Database
(NED) for any other galaxies within a projected radius of 1 Mpc
($\approx 20$ arcmin) and $\pm 1000$ km s$^{-1}$ in velocity of
3C\,305, and found only three for which NED has redshifts, of which
the nearest (and brightest) is the spiral MCG +11-18-009, 130 kpc away
in projection and 1.5 magnitudes fainter; the remaining two are so far
away that it seems unlikely they are physically associated with IC
1065. It thus certainly seems plausible that 3C\,305's host, though a
massive galaxy (H82) lies in a very poor environment, explaining the
lack of group-scale X-ray emission: similar
conclusions based on targeted optical spectroscopy were reached by
Miller \etal\ (2002). 3C\,305 thus differs from FRI radio galaxies of
comparable radio power (e.g. 3C\,66B, 3C\,442A, M87, 3C\,465) which
lie in rich groups or clusters (Croston \etal\ 2003; Hardcastle
\etal\ 2007; B\"ohringer \etal\ 1995; Hardcastle, Sakelliou \& Worrall
2005) and also from at least some low-power FRII sources (e.g.
3C\,285, which lies in a poor to moderate group: Hardcastle
\etal\ 2007).

There is also no evidence for any emission spatially associated with
compact features of the radio galaxy such as the jets or hotspots.
Synchrotron emission is known from a large number of hotspots and jets
in low-power radio galaxies of this type, but none is apparent in 3C\,305.

\subsection{Spectral fits}
\label{regions}

Spectra were extracted, in the manner described in Section
  \ref{astrometry}, for the following regions:
\begin{enumerate}
\item `Core': a 1.4-arcsec circle centred on 14:49:21.618, +63:16:13.90, which
  is the approximate centroid of the hard X-ray emission discussed
  above and also lies just to the SW of the base of the radio jet.
\item `Whole source': a 15-arcsec circle with the same centre,
  encompassing all of the emission plausibly associated with the
  source.
\item `Extended': the whole source minus the core region.
\item `Hotspot': the bright region of X-ray emission around and to the
  NE of the NE radio hotspot, defined as an elliptical region with
  semi-major axis 2.3 arcsec, semi-minor axis 1.6 arcsec, and position
  angle 35$^\circ$ (north through east) centred at 14:49:21.937,
  +63:16:15.29: this region was selected because it is the brightest
  sub-region of the X-ray emission and is coincident with the clearest
  depolarization silhouette. (Note that, while roughly coincident
    with the radio hotspot, this region is much larger than the
    hotspot and is not, as noted above, associated with any detected
    non-thermal emission.)
\item `Wings': the extended emission elongated roughly perpendicular to
  the jet axis across the centre of the source, enclosed by an
  elliptical region with semi-major axis 3.8 arcsec, semi-minor axes
  1.3 arcsec, and excluding the core region.
\end{enumerate}

\begin{figure}
\epsfxsize 8.5cm
\epsfbox{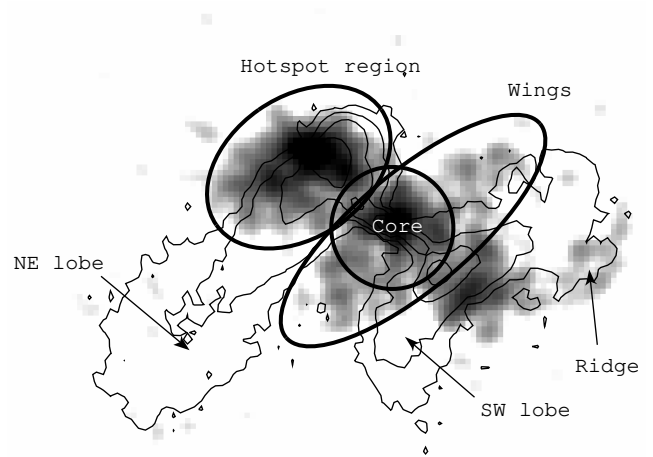}
\caption{Locations of the core and hotspot X-ray extraction regions.
  The `Whole source' region encompasses all the emission seen here.
  Grayscale shows the X-ray emission binned and smoothed as in Fig.\ \ref{chandra-overlay}.
  Contours are from the 8.4-GHz radio map at $3\sigma \times
  (1,4,16\dots)$ mJy beam$^{-1}$.}
\label{bwfinding}
\end{figure}

The locations of the core, hotspot and wings regions are shown in
Fig.\ \ref{bwfinding}. For all these regions the background was taken
to be an annulus between 15 and 20 arcsec concentric with the core,
which contains no detected point-like or extended emission. We did not
attempt to take local background for sub-regions of the source because
of the non-uniformity of the extended emission.

\begin{figure}
\includegraphics[width=.33\textwidth, angle=270]{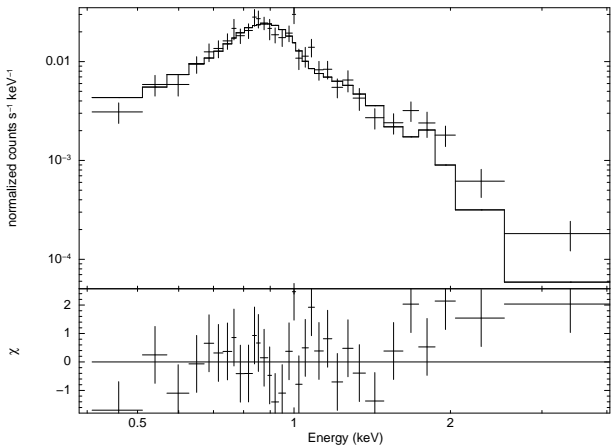}
\caption{X-ray count spectrum, fitted model and contributions to
  $\chi^2$ for the `Extended' region discussed in the text.}
\label{spectrum}
\end{figure}

Simple power-law models were not an acceptable fit to any of the
extracted spectra (the lowest reduced $\chi^2$ value, 3.4, was found
in the core region). All of the spectra had the characteristic excess
between 0.7 and 1.1 keV that is typical of emission from
keV-temperature partially ionized gas. Accordingly, we next fitted
with single Astrophysical Plasma Emission Code (APEC) models with free
abundance relative to solar and temperature; these
models accurately represent the continuum and line emission from a
collisionally ionized plasma\footnote{See http://atomdb.org/}. Results
of these fits are given in Table \ref{fit-results} and an example
  X-ray spectrum is shown in Fig.\ \ref{spectrum}.

\begin{table*}
\caption{Net counts from, and APEC fits to the spectra of various
  regions of 3C\,305 and the corresponding bolometric X-ray
    luminosities of the thermal component.}
\label{fit-results}
\begin{tabular}{lrrrrr}
\hline
Region&Net counts&$\chi^2/{\rm dof}$&$kT$&Abundance&$L_{\rm Bol,\ unabs}$\\
&(0.5-5.0 keV)&&(keV)&(solar)&($\times 10^{41}$ erg s$^{-1}$)\\
\hline
Core&$255 \pm 16$&31.3/8&1.02&0.10&--\\
Whole source&$1091 \pm 35$&78.2/41&$0.81\pm 0.02$&$0.15 \pm 0.02$&$4.8 \pm 0.5$\\
Extended&$836 \pm 32$&40.6/33&$0.78 \pm 0.02$&$0.15 \pm 0.03$&$3.9 \pm 0.4$\\
Hotspot&$384 \pm 20$&19.4/13&$0.80 \pm 0.03$&$0.20 \pm 0.05$&$1.7 \pm 0.3$\\
Wings&$112\pm 11$&6.1/5&$0.97 \pm 0.10$&$0.12_{-0.05}^{+0.08}$&$0.6 \pm 0.1$\\
\hline
\end{tabular}
\end{table*}

From this table it can be seen that the single-APEC fits to the
extended region, wings and hotspot are acceptable but the fit to the core is
very clearly not, and the fit to the whole source is similarly rather
poor (reduced $\chi^2 \sim 2$). This is not unexpected if the hard
counts noted above come from the location of an absorbed AGN. We
therefore fitted a model to the core spectrum consisting of an APEC
plus an intrinsically absorbed power law. The absorbing column, photon
index and normalization of the power law were free to vary: the
abundance of the APEC component was set to 0.2 solar, based on what is
found in the free-abundance fits, to reduce the
number of free parameters. This model was a good fit ($\chi^2/dof =
1.7/6$) with $kT = 0.73 \pm 0.06$, photon index $\Gamma = 1.7\pm 0.5$ and intrinsic absorption column
$3_{-2}^{+5} \times 10^{21}$ cm$^{-2}$. The photon index is consistent
with what is seen in the absorbed components of narrow-line radio
galaxies (e.g.\ Hardcastle, Evans \& Croston 2009). (The implications
of this model for the X-ray emission of the AGN are discussed in
Section \ref{core}.)

We then took this best-fitting model for the power-law
component of the core, with all parameters frozen, and added it to the
APEC model fitted to the whole source; we found that, as might be
expected, the fit was much improved ($\chi^2/dof = 52.8/41$, $kT = 0.78
\pm 0.02$, abundance $0.18 \pm 0.03$). Thus it is reasonably clear
that the extended emission can adequately be fitted with a model in
which it originates in a single-temperature, thermal, collisionally
ionized plasma (though our data are not good enough to rule out
multi-temperature models). The unabsorbed luminosity in the thermal
component of this X-ray emission
is $1.8 \times 10^{40}$ erg s$^{-1}$ (2-10 keV rest-frame) or $4.4
\times 10^{41}$ erg s$^{-1}$ (bolometric). This is much lower than the
estimated line luminosity of the system, $\sim 10^{44}$ erg s$^{-1}$
(H82); we return to this point below (Section \ref{energetics}).

The low best-fitting abundances in the APEC fits are noteworthy.
Abundances in the hot phase of the ISM/IGM in groups of galaxies,
which are the most obvious comparison to 3C\,305, tend to be
significantly higher than our best-fitting values, particularly at the
centres of the host groups (e.g. Johnson \etal\ 2011). There are
several possible explanations for our results. One is that this is a
real effect: it might, for example, arise if the hot gas comes from a
recent interaction between the radio galaxy and low-abundance cold
gas, as we will argue in Section \ref{conclusions}. However, low
abundances can also be an artefact of fitting the wrong model: if a
single-temperature model is fitted to a two-temperature plasma, the
abundance is systematically low (Buote \& Fabian 1998) while a similar
effect might be expected if the plasma is photoionized rather than
collisionally ionized (see Section \ref{ionization}). We cannot rule
out these explanations from the X-ray spectroscopy alone.

Finally, we note that the consistency between the APEC parameters
estimated for the hotspot, wings and the whole extended region, and
the good fits obtained with Galactic $N_{\rm H}$ values, suggests that
there is no significant effect of spatially variable intrinsic X-ray
absorption on the X-ray spectrum. The deep, narrow component of H{\sc
  i} ($N_{\rm \matHI} = 5.4 \times 10^{20}$ cm$^{-2})$ observed by
Morganti \etal\ (2005a) is in front of the SW lobe, and there is
relatively little X-ray emission there (Fig.\ \ref{chandra-overlay}),
though the statistics are not good enough to say whether this is an
effect of absorption. The broad, shallow component corresponds to a
higher column density ($N_{\rm \matHI} = 2 \times 10^{21}$ cm$^{-2}$
for an assumed spin temperature of 1000 K) and Morganti
\etal\ place this in front of the NE radio lobe, where the bright
`hotspot' X-ray emission is seen. Fitting an APEC model with both
Galactic and intrinsic absorption to the spectrum from this region, we
can place a 99\% confidence upper limit on the intrinsic column
density of $<1.5 \times 10^{21}$ cm$^{-2}$ if the temperature and
abundance are allowed to vary, and $<1.1 \times 10^{21}$ cm$^{-2}$ if
they are fixed to force consistency with the overall extended
spectrum. It seems most likely, therefore, that at at least some and
quite possibly all of the H{\sc i} is behind, or mixed in with, some
or all of the X-ray emitting gas. In practice, the neutral hydrogen is
unlikely to be cleanly separated from the X-ray-emitting plasma: it is
more likely that the two form part of a multi-phase medium, together
with the warm gas emitting in optical emission lines, with the H{\sc
  i} having a comparatively low filling factor (as suggested by O'Dea
\etal\ 1994). In this situation, we would expect a much lower
effective absorbing column in X-ray spectroscopy than is observed
toward the radio continuum. We also note that a higher spin
  temperature than assumed by Morganti \etal\ (2005a), which is
  perfectly possible in the radiative environment of the AGN (e.g.
  Liszt 2001; Holt \etal\ 2006) would increase the discrepancy between
  the X-ray limit and H{\sc i} measurements, and would therefore
  strengthen the argument in favour of a complex multi-phase medium.

\subsection{Hardness ratio analysis}

An alternative method of searching for any spectral variation as a
function of position in the source is to consider hardness ratios.
Generically, these are based on the ratio of X-ray fluxes in two bands
(`hard' and `soft'). Their advantages are that they can be calculated
in regions where there are too few counts to fit X-ray spectra, and
that their interpretation does not pre-suppose any physical model. In
general, estimating hardness ratios as a function of position for an
extended X-ray source requires some idea of the effective area as a
function of position --- the flux corresponding to a given number of
counts is not constant over the detector. Given the small physical
size of 3C\,305, though, we can make the simplifying assumption that
the energy-dependent effective area is constant over our region of
interest. We divide the counts from the source into 3 energy bands,
`soft' (0.5 - 1.0 keV), `medium' (1.0 - 2.0 keV) and `hard' (2.0 - 7.0
keV). Then we define hardness ratios as follows:
\[
H_1 = 1 + \frac{C_M - C_S}{C_M + C_S}
\]
\[
H_2 = 1 + \frac{C_H - C_M}{C_H + C_M}
\]
where $C_S$, $C_M$ and $C_H$ are the counts in a given region in the
three bands. As the errors on the counts are Poissonian, the errors on
these hardness ratios have a distribution that is hard to describe
analytically. See Appendix A for a discussion of the approach we adopt.

\begin{figure}
\epsfxsize 8.5cm
\epsfbox{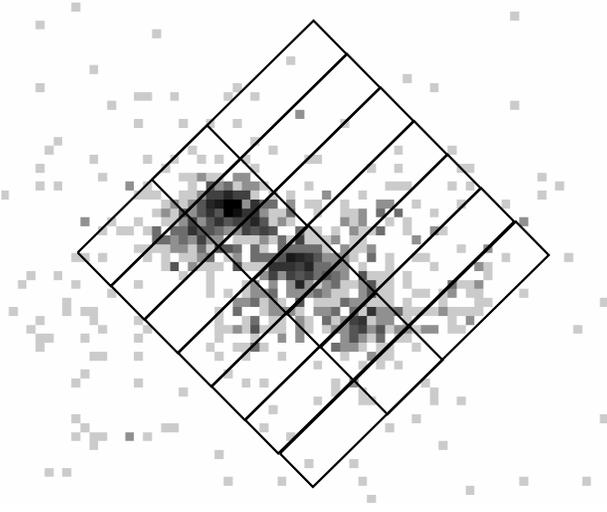}
\caption{Regions used for hardness ratio extraction}
\label{hr-regions}
\end{figure}

\begin{figure*}
\epsfxsize 8.5cm
\epsfbox{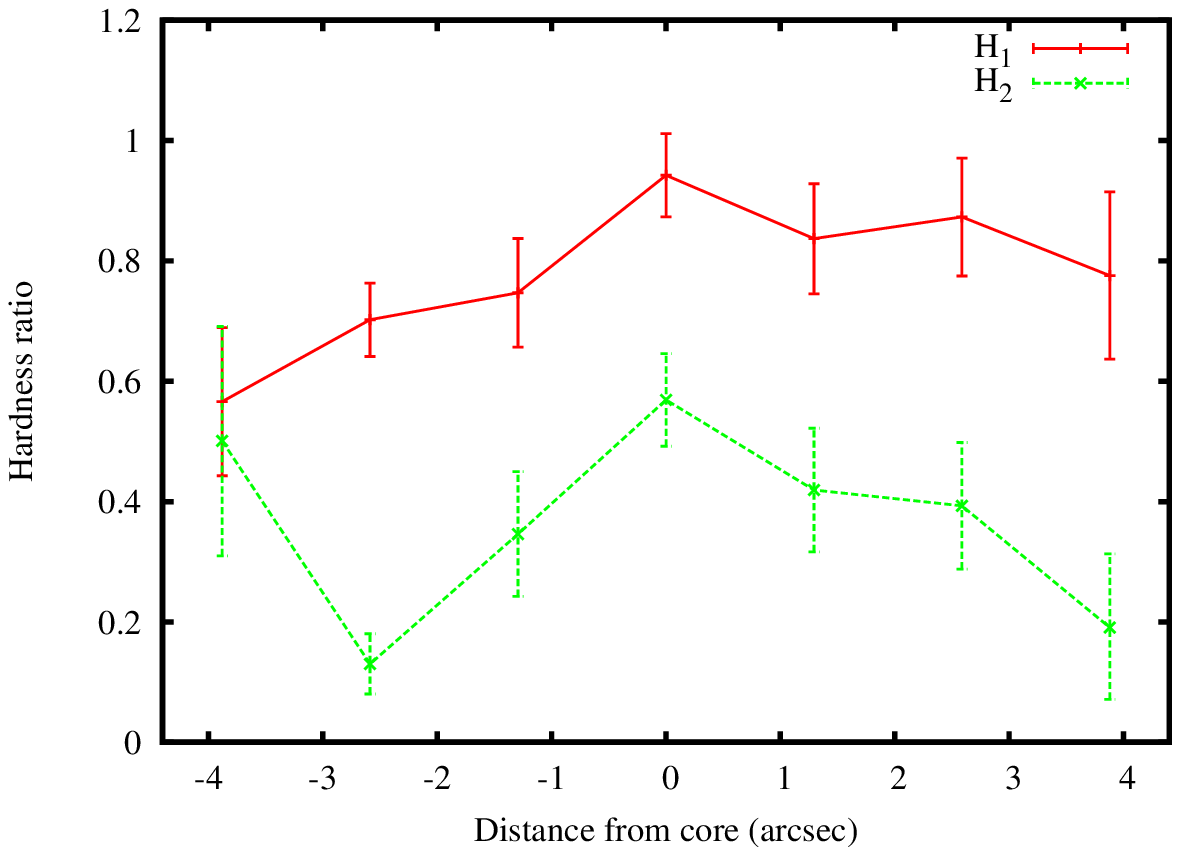}
\epsfxsize 8.5cm
\epsfbox{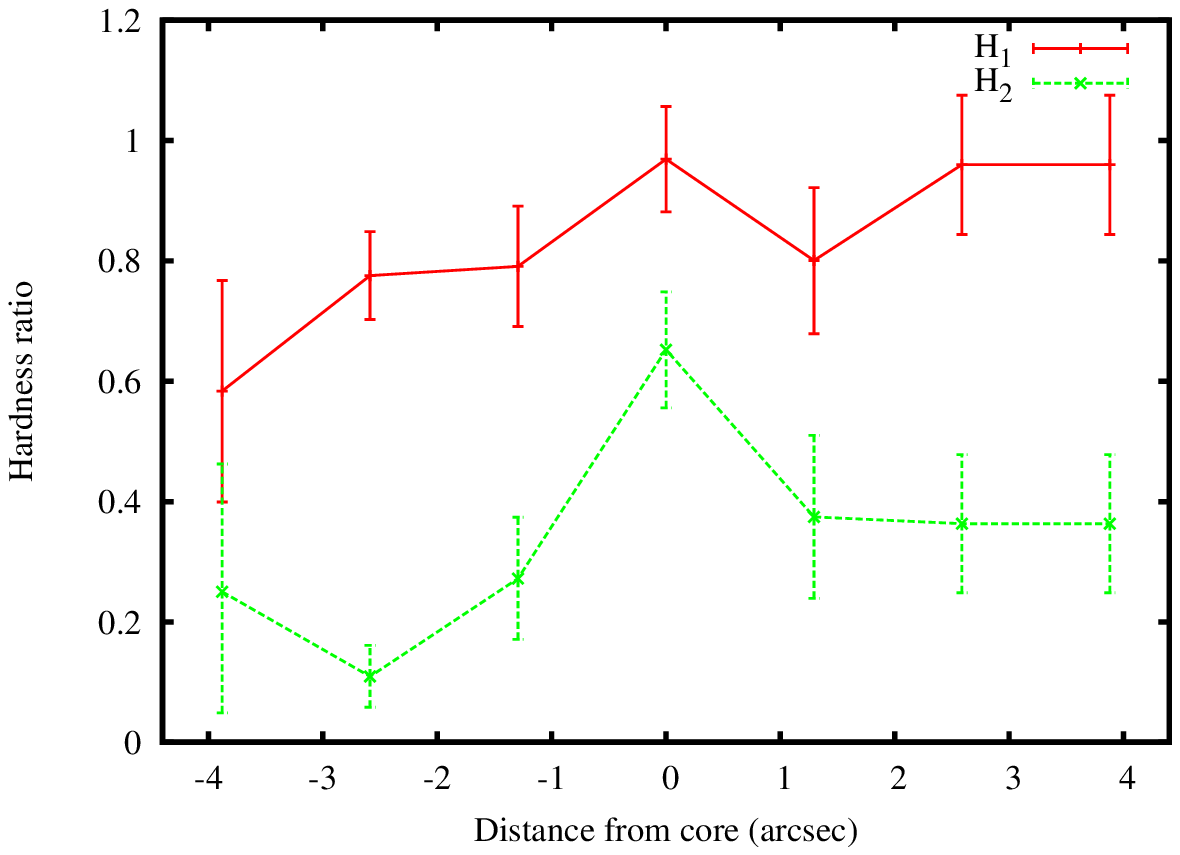}
\caption{Bayesian estimates (see Appendix A) of the hardness ratios,
  as defined in the text, as a function of position for (left) large
  and (right) small bins positioned along the jet axis as shown in
  Fig.\ \ref{hr-regions}. Distances are defined along the jet axis,
  relative to the core, and increase in the direction from NE to SW.}
\label{hr-results}
\end{figure*}

Rather than replicate the regions used for our spectral analysis, we
decided to divide the source up into identically sized rectangular
regions along the jet axis. This allowed us to search for spectral
variation as a function of distance from the nucleus, which might be
expected in photoionization models for the extended X-ray emission.
The minimum size of these regions was determined by the requirement
that we should obtain enough counts in the hard band to allow a
calculation of the hardness ratio. This allowed 7 regions along the
jet axis. We centred the middle region on the core, while two of the
outer regions more or less coincide with the positions of the radio
hotspots. We defined large and small versions of our regions: the
large versions contain essentially all the emission from the source,
while the small versions are restricted to $\pm 1.05$ arcsec around the
jet axis (Fig.\ \ref{hr-regions}). We then estimated the hardness
ratios $H_1$ and $H_2$ in each bin using the method described in
Appendix A. The background for this bin size
was $<1$ count in each bin and so we neglected it. Results are plotted
in Fig.\ \ref{hr-results}.

We see that there is little strong evidence of variation of hardness
ratio as a function of position, consistent with the spectral analysis
presented above. We note first of all that the hardness ratios of both
sets of regions are consistent within the joint errors everywhere
(this is not surprising, as they are not independent, but it gives an
indication that there is no significant difference between the
off-axis and on-axis parts of the source). There is a clear central
peak in $H_2$ in both sets of plots, which is not surprising given the
overdensity of hard counts in the core region discussed in Section
\ref{astrometry}. There also seems to be a marginally significant
deficit of hard counts in the region at $\sim -3$ arcsec from the
core, which corresponds to the peak of the `hotspot' region in our
spectral analysis. But other than that, there seems to be no clear
evidence for HR variation as a function of position: within the
errors, $H_1$ is consistent with being constant ($\sim 0.8$)
throughout the source, while $H_2 \sim 0.3$. This supports our
conclusion from spectral analysis that the bulk of the extended
emission is adequately represented by a model with a single X-ray spectrum
and that there are no strong variations of absorbing
column with position in the source.

\section{Discussion}

\subsection{The X-ray core and the state of the AGN}
\label{core}

As noted above, the best-fitting absorbed power-law model for the core
region gives a spectrum which has properties reasonably similar to
those of the nuclei of other narrow-line radio galaxies. 

However, the unabsorbed 2-10 keV luminosity of the power-law component
in this model is only $(5\pm 1) \times 10^{40}$ erg s$^{-1}$ (where
the error is derived from the $1\sigma$ error on normalization only),
which is extremely low compared to most 3CRR radio galaxies, and would
place 3C\,305 an order of magnitude or more below the correlation
between radio power and X-ray nuclear power observed for narrow-line
radio galaxies (Hardcastle \etal\ 2009). Given that few NLRG have a
column density $<10^{22}$ cm$^{-2}$ towards the heavily absorbed
component, it may be that we are seeing absorption towards only the
jet-related component, and that the accretion disc is behind a much
larger column, although we note that, since there is only an upper
limit on the compact radio core flux, we would not expect a strong
jet-related component either (see Hardcastle \etal\ 2009 and
references therein). Jackson \etal\ (2003) suggest that the optical
nuclear component is consistent with being a reddened quasar with $A_V
> 4$, but this only requires $N_H \ga 7 \times 10^{21}$ cm$^{-2}$,
consistent with our best-fitting model. However, the estimate given by
Jackson \etal\ of the $M_V$ of the hidden quasar, $-22.5$, would lead
us to expect a much more luminous X-ray source (e.g. Elvis
\etal\ 1994), requiring a much higher obscuring column. The {\it
  Spitzer} 24-$\mu$m flux quoted by Dicken \etal\ (2010) implies that
$L_{\rm IR} \sim 2 \times 10^{43}$ erg s$^{-1}$, which, on the
correlation between $L_{\rm IR}$ and accretion-related X-ray reported
by Hardcastle \etal\ (2009), would lead us to expect a nuclear X-ray
luminosity $\sim 10^{43}$ erg s$^{-1}$. Finally, we could
  estimate the nuclear X-ray luminosity from the [O{\sc iii}]
  luminosity, using the correlation of Heckman \etal\ (2005), but this
  relies on the uncertain reddening correction, as well as on the
  excitation mechanism for the [O{\sc iii}]; if we do not correct
  the H82 [O{\sc iii}] fluxes for reddening (see the discussion of
  this point in Section \ref{energetics}), we would again obtain an
  expected 2-10 keV luminosity around $10^{43}$ erg s$^{-1}$, but this
  could be substantially lower if the emission-line regions are
  partially or wholly shock-excited, or substantially higher if the
  reddening correction is large, making such estimates very uncertain.

All these data are difficult to reconcile. One
obvious explanation is that 
the source has varied substantially on a timescale of years to
decades, although there is no evidence for this in our multi-epoch X-ray
data or in the {\it XMM-Newton} data studied by Evans \etal\ (2008).
There would be a significant time lag between a substantial drop in
the optical/X-ray output of the accretion disc and the corresponding
drop in the luminosity of re-radiated emission from the `torus'
region, which could help to explain the difference between the 
X-ray and IR luminosities. Alternatively, it may be that the AGN is
Compton-thick, and that the intrinsic X-ray luminosity currently is
roughly the $\sim 10^{43}$ erg
  s$^{-1}$ implied by the mid-IR data.

\subsection{Ionization mechanism}
\label{ionization}

M09 raised the important question: what is the ionization mechanism of
the X-ray-emitting gas? Although we obtain good fits with a thermal
(APEC) model, implying consistency with a collisional-ionization
scenario, we cannot rule out photoionization directly from the {\it
  Chandra} spectra. M09 pointed out that the ratio between the [O{\sc
    iii}] flux and the soft X-ray flux was of order unity (compare our
bolometric X-ray luminosity of $4.4 \times 10^{41}$ erg s$^{-1}$ with
the [O{\sc iii}] luminosity, calculated from the flux quoted by M09,
of $1.5 \times 10^{41}$ erg s$^{-1}$), which is often observed in
Seyfert galaxies, where the X-ray-emitting medium is thought to be
directly photoionized by the AGN (e.g.\ Bianchi \etal\ 2006). However,
in 3C\,305, there is reasonably strong circumstantial evidence that
the emission lines are shock-ionized, as discussed by H82, and also
some direct evidence for this (the detection of [Fe{\sc ii}] emission
at the likely location of the strongest shocks, Jackson \etal\ 2003).
Any agreement between the luminosities in X-rays and [O{\sc iii}] may
therefore be coincidental, or at least be a reflection of the fact
that the same jet is driving both. There are also some practical
difficulties with a photoionization model for the X-rays, including
the fact that we see no direct evidence in the X-ray for a luminous
AGN at all (Section \ref{core}) let alone one with the luminosity
needed to drive the observed extended X-ray emission. Perhaps the
  strongest argument against photoionization is morphological: in
  addition to the fact that the apparent association between the radio
  source and the X-ray emission would have to be a coincidence in such
  a model, several features of the extended X-rays not seen in the
  data of M09, such as the `wings' in the centre of the source and the
  ridge at the edge of the SW lobe, are very hard to reconcile with
  the expected roughly conical morphology for photoionization, and yet
  have X-ray spectra and hardness ratios completely consistent with
  those of the rest of the source.

To investigate a photoionization model quantitatively we follow the
analysis in H10, which was itself based on the analysis of X-ray
photoionization in a Seyfert galaxy by Weaver \etal\ (1995). Consider
the entire `hotspot' region to be photoionized by a hidden AGN and,
for simplicity (and also because it minimizes the distance of the
hotspot from the AGN and so makes the energetic requirements
as low as possible) suppose that the X-ray structure lies in the plane
of the sky. Then the `hotspot' region can be modelled as a uniformly
filled cone between two radii, centered on the AGN, of $r_{\rm in}$
and $r_{\rm out}$, which we measure to be roughly 2.6 and 6.4 kpc
respectively. The opening angle of the cone is 67 degrees, which means
that we need to take the curvature of the volume elements into
account, so our analysis is slightly different from that of H10. The
surface brightness of the hotspot region is roughly constant, with no
sign of a strong gradient away from the nucleus, so let
the density be given by $n(r) = n_{\rm in} (r/r_{\rm in})^{-2}$, so as
to keep the ionization parameter constant: then
\begin{equation}
L_{\rm line} = \int_{r_{\rm in}}^{r_{\rm out}} \Omega r^2 n^2(r)
j(\xi) {\rm d}r
\end{equation}
where $\Omega$ is the solid angle subtended by the cone ($\Omega =
2\pi (1-\cos \theta)$ where $\theta$ is the half-opening angle: thus
in this case $\Omega = 1.04$ sr) and $j(\xi)$ is the volume emissivity
for a given ionization parameter. This gives us
\begin{equation}
L_{\rm line} = \Omega n_{\rm in}^2 r_{\rm in}^4 j(\xi) \left
[\frac{1}{r_{\rm in}} - \frac{1}{r_{\rm out}}\right ]
\end{equation}
and so for given values of $r_{\rm in}$, $r_{\rm out}$ and $\Omega$,
and for a known value of $j(\xi)$ (we follow Weaver \etal\ and H10 and
use $j(\xi) = 10^{-24}$ ergs cm$^{-3}$ s$^{-1}$) the luminosity
determines the density throughout the photoionized region. For the
observed 0.1-10 keV luminosity of the hotspot region, $1.4 \times
10^{41}$ erg s$^{-1}$, we have $n(r_{\rm in}) = 0.66$ cm$^{-3}$ and
$n(r_{\rm out}) = 0.11$ cm$^{-3}$. These densities are not in themselves
  particularly implausible -- they are roughly comparable to the
  densities estimated, obviously on quite different assumptions, from
  the APEC model in Section \ref{pr-balance}, and are similar to the
  results we obtained for the photoionization analysis of 3C\,171. However, the
requirement that the ionization parameter $\xi = L_{\rm I}/nr^2 > 100$
erg s$^{-1}$ cm (Weaver \etal 1995) then implies an ionizing
luminosity $L_{\rm I} > 4 \times 10^{45}$ erg s$^{-1}$. Even if we
relax the required ionization parameter by an order of magnitude, we
require $L_{\rm I} > 4 \times 10^{44}$ erg s$^{-1}$.

Can this luminosity be supplied by the AGN? As noted above (Section
\ref{core}) our best estimate of the unabsorbed X-ray luminosity of
the nucleus is only $5 \times 10^{40}$ erg s$^{-1}$: no correction
factor for the non-X-ray part of the ionizing continuum (a factor of a
few) can make up this discrepancy. However, as further noted above,
the X-ray luminosity seems anomalously low for a NLRG given the radio
luminosity of the source and the mid-IR luminosity of the nuclear
region. If the AGN has the much higher nuclear X-ray luminosity ($L_X
\sim 10^{43}$ erg s$^{-1}$) implied by the mid-IR data, or has had it
in the recent past, then the photoionization model is closer to being
viable, but still an order of magnitude off even for the very
favourable assumptions that we have used. Although our model
parameters are approximate (as we noted above, the extended X-ray
emission is not very convincingly modelled as a uniformly filled cone)
there are no obvious changes that we could make that would produce a
very low density and thus reduce the luminosity needed to produce the
ionization parameter required for X-ray emission. As in the case of
3C\,171, we conclude that a photoionization model for the extended
X-ray emission is difficult to sustain, and so in what follows we make
use of densities derived from the APEC fits to the X-ray emission.

\subsection{Physical conditions in the depolarizing medium}
\label{pr-balance}

From the normalization of the APEC fit to the `hotspot' region
described in Section \ref{regions}, we can estimate the density of
X-ray emitting plasma, on the assumption that the hotspot region is a
uniformly filled prolate ellipsoid with the dimensions we used for the
extraction. This gives an electron density of $(1.9 \pm 0.1) \times
10^{-1}$ cm$^{-3}$ and a pressure of $(4.5 \pm 0.3) \times 10^{-11}$
Pa, or $(4.5 \pm 0.3) \times 10^{-10}$ dyn cm$^{-2}$. H82 estimate
pressures in the emission-line regions in the range $10^{-9}$ --
$10^{-10.5}$ Pa, which is rather higher than what we find here, but
comparable at the lower end, so that it is possible that the
X-ray-emitting plasma is in rough pressure balance with the
line-emitting material. Neither the density nor the temperature of the
line-emitting material is well constrained. The minimum energy density
in the lobes $U_{\rm min}$ is of order $3 \times 10^{-11}$ J m$^{-3}$
($3 \times 10^{-10}$ erg cm$^{-3}$), so assuming a fully tangled field
the lobe pressures ($p_{\rm min} = U_{\rm min}/3$) are significantly
less than the pressures in the hotspot region, which may help to
explain why the lobe material is driven away from the jet axis.
(Consistent with this, the pressure in the `wings' region, if we
assume it to be a uniformly filled oblate ellipsoid with the
dimensions of the extraction region, is $(1.7 \pm 0.4) \times
10^{-11}$ Pa, very similar within the errors to the minimum lobe
pressure; thus it is possible that the `wings' are shocked material
that has been driven back towards the centre by the expanding radio
lobes.)

The `hotspot' region is presumably responsible for the depolarization
silhouette seen in front of the NE hotspot in the radio. Following
Hardcastle (2003), we expect the degree of polarization at a
given frequency $\nu$ to be given by
\begin{equation}
p_\nu = p_i \exp(-C\nu^{-4})
\label{depoleq}
\end{equation}
where $p_i$ is the intrinsic degree of polarization and $C$ depends on
the physical conditions in the depolarizing medium:
\begin{equation}
C = 2K^2(n_e B_\parallel)_f^2 d R c^4
\end{equation}
where $K$ is a constant with value $8.12 \times 10^{-3}$ rad nT$^{-1}$
m kpc$^{-1}$, $(n_e B_\parallel)_f^2$ is the dispersion in the product
of number density and magnetic field strength along the line of sight
in units of nT m$^{-3}$, $d$ is the size scale in kpc of the
individual depolarizing regions (regions of uniform magnetic field)
and $R$ is the line-of-sight depth through the medium in kpc. For the
depolarization silhouette we obtain average values of $p_\nu$ by
integrating polarized and total intensity over the most depolarized
region, and we can then solve eq.\ \ref{depoleq} for $C$, obtaining $C
= 8.5 \times 10^{39}$ Hz$^4$. We take $R = 1.3$ kpc (the semi-minor
axis of the ellipsoidal region discussed above). We know that $d$ must
be much less than the resolution at X-band (8.6 GHz), $d\ll 0.16$ kpc, and as we
know there are still some unpolarized regions even at the full K-band
resolution, it seems likely that $d< 70$ pc: $d = 70$ pc should give
us a limit. If we further assume that the electron density is constant
and that the Faraday dispersion comes only from field reversals, then
$(B_\parallel)_f^2 \approx B^2$, and this tells us that $B \ga 1.6$ nT
(16 $\mu$G).
Following H10, we expect an upper limit to be such that the energy
density in the magnetic field is less than that in the gas, i.e.
\begin{equation}
\frac{3}{2}nkT > \frac{B^2}{2\mu_0}
\end{equation}
and for the densities and temperatures we have determined above, this
implies that $B<10$ nT (100 $\mu$G), which is consistent with the
lower limit derived from depolarization, and requires that $d>2$ pc.
We conclude from this that it is very plausible that the X-ray
emitting plasma is the depolarizing medium, and that it is magnetized
with a magnetic field energy density which is within an order of
magnitude or so of the
thermal energy density. If the X-ray-emitting material is the
depolarizing medium, its filling factor must be close to unity.
As in 3C\,171, the self-consistency of these calculations gives
  some support to a model in which the gas is collisionally ionized
  rather than photoionized.

\subsection{Source energetics}
\label{energetics}

We now have what seems likely to be a full picture of the various
phases of gas known to be associated with the outflow in 3C\,305. We
emphasise that we have no direct evidence for outflow in the
X-ray-emitting component, but it is very hard to imagine a model in
which this component is not moving with the optical line-emitting gas,
which is morphologically similar (Section \ref{environment}) and in
rough pressure balance (Section \ref{pr-balance}). In what follows we
assume that the X-ray, optical line and H{\sc i}-emitting material are
multiple phases of a single outflow. Guillard \etal\ (2012) have
recently shown that the kinematics of the {\it molecular} hydrogen in
3C\,305 (and in several other sources) do not appear to be consistent
with those of the H{\sc i} or the optical emission-line gas, although
we note that some of the energy supplied by the jet is likely to go
into excitation of the molecular gas. It is possible that the {\it
  Spitzer} observations of Guillard \etal\ did not have the
sensitivity to detect rapidly outflowing material; alternatively, it
is possible that the molecular gas is actually dissociated by the
process that drives the fast outflow, and so shows apparently
different kinematics for this reason. In any event, as the molecular
material presents what appears to be a somewhat different kinematic
behaviour, we neglect this phase in what follows.

In order to calculate the contributions of each phase to the energy
budget of the source, we need to know the speed of the outflow. H82
estimated the kinetic energy in the line-emitting gas on the assumption
that its speed is the maximum radial velocity seen in the emission
lines ($v_r = 260$ km s$^{-1}$). This is a lower limit, since the
jets, which presumably drive the outflow, are not pointing directly
towards us, and in fact given their two-sidedness are probably
reasonably close to the plane of the sky, so that $v_r$ is probably
significantly less than $v$. If the splitting and
  broadening of the optical emission lines near the NE lobe (Morganti \etal\ 2005a) is
  due to us observing shocked material on both sides of the source,
  then the true outflow velocity would have to be {\it at least} half
  the difference between the centroids of the two velocity components,
  which is $\sim 400$ km s$^{-1}$, consistent with the above.
Moreover, in some models of the
acceleration of emission-line clouds by shocked gas, it is possible
for the clouds to be moving significantly slower than the bow shock
(O'Dea \etal\ 2002). However, $v_r$ and the similar velocities
estimated from the H{\sc i} observations (Morganti \etal\ 2005a) are
the only {\it direct} constraints on velocity that we have.

\begin{table*}
\caption{Contributions to the energetics of the outflow in 3C\,305}
\label{energies}
\begin{center}
\begin{tabular}{lrrrrr}
\hline
Phase&Mass&Ref.&Internal energy&Kinetic energy&Contribution to total\\
&($M_\odot$)&&(erg)&(erg)&(erg)\\
\hline
Hot gas&$7 \times 10^7$&1&$6 \times 10^{56}$&$3\times 10^{56}$&$6
\times 10^{56}$\\
Emission-line gas&$4 \times 10^5$&2&--&$2 \times 10^{54}$&$2 \times 10^{54}$\\
Cold gas&$10^7$&2&--&$4 \times 10^{55}$&$4 \times 10^{55}$\\
Magnetic field in hot gas&--&1&$>10^{55}$&--&$10^{55}$\\
Lobes&--&1&$>3 \times 10^{56}$&--&$3 \times 10^{56}$\\
Total&&&&&$10^{57}$\\
\hline
\end{tabular}
\end{center}
\medskip
A dash indicates that the quantity in question is negligible.
References for gas masses are as follows: (1) This paper; (2) Morganti
\etal\ (2005a). Kinetic energies are calculated on the assumption that
$v = 650$ km s$^{-1}$ (see the text). Half the internal energy of the
hot phase is assumed to be pre-existing (i.e. it was the internal
energy of the unshocked gas) and so does not count towards
the total energy input from the radio source.
\end{table*}

One approach to determining the true velocities is to estimate the
expansion speed from the observed X-ray and radio properties,
following H10. If we think
the X-ray emission represents shocked external thermal gas, then the pressure in the `hotspot' region is a proxy of the post-shock
pressure; the minimum pressure in the large-scale lobes must be equal
to or greater than the pressure in the large-scale unshocked gas in
which they are embedded. We can then
apply the shock pressure relation
\def\mach{{\cal M}}
\begin{equation}
{{p_2}\over p_1} = {{2\Gamma{\mach}_1^2 + (1-\Gamma)}\over{\Gamma +1}}
\end{equation}
where $\Gamma$ is the adiabatic index, 5/3 in this case, with $p_2 =
4.5 \times 10^{-11}$ Pa (the pressure in the `hotspot' region) and $p_1 =
1 \times 10^{-11}$ Pa (the minimum pressure in the lobes), to find that
$\mach_1 \sim 2$. Since we also know that $T_2 = 0.8$ keV, and
\begin{equation}
{{T_2}\over T_1} = {{\left[2\Gamma\mach_1^2 +
(1-\Gamma)\right]\left[\Gamma - 1 + 2/\mach_1^2\right]}\over
\left(\Gamma+1\right)^2}
\label{temp}
\end{equation}
we find that the unshocked temperature $T_1$ is 0.4 keV, the sound
speed in the medium is 330 km s$^{-1}$, and the shock speed $v_{\rm
  shock}$ would be 650 km s$^{-1}$. The temperature $T_1$ is certainly
plausible for the hot halo of an isolated massive galaxy (see Section
\ref{environment}) and the
  expected undisturbed bolometric X-ray luminosity of such an object, $\sim
  10^{41}$ erg s$^{-1}$, is not ruled out by our limits on the
  large-scale emission. If we assume that the emission-line material
  is moving with the shock, the speed would be consistent with the observed
radial velocities if the angle to the plane of the sky $\theta$ were
$24^\circ$ ($v_r = v_{\rm shock} \sin \theta$), which seems reasonable
(and is consistent with the estimates of M09 from jet sidedness); if
the emission-line clumps are actually slower than the shock speed,
this gives a lower limit on $\theta$.
Kinetic energies for the outflowing material are a factor
$\csc^2(24^\circ) \approx 6$ higher if we assume that the actual
outflow speed $v = v_{\rm shock}$
than if we assume $v = v_r$, and we use this speed in what follows.

Morganti \etal\ (2005a) estimated a lower limit on the mass of the
emission-line material actually associated with the outflow, for one
lobe, of $>2 \times 10^5 M_\odot$; we adopt twice this value to
characterize the overall outflow, but this low mass still makes the
kinetic energy contribution of this phase quite negligible. We derive
a mass estimate of the X-ray-emitting gas from the normalization of
the APEC fits to the `hotspot' region, which gives a density as
discussed in Section \ref{pr-balance}. If we very roughly assume a
uniform density for the X-ray-emitting gas throughout the source, then
scaling the mass for the north hotspot region up to the total by the
ratio of the emission measures determined from spectral fitting, we
estimate that the mass of the X-ray-emitting gas is $7 \times
10^7M_\odot$ -- the systematic error on this estimate is of
  course large, principally because we do not know the geometry of the
  emitting region, but is unlikely to exceed a factor $\sim 2$.
Morganti \etal\ (2005a) argue that the mass of H{\sc i} seen in the
system is $\sim 10^7 M_\odot$, so that it dominates the
  energetics of the cold phase of the outflow. Taking all these three
phases together, the total mass of gas in the outflow
approaches $10^8 M_\odot$, and the kinetic energy is at least $6
\times 10^{55}$ erg ($v=v_r$), and is $\sim 3.5 \times 10^{56}$ erg if
$v = v_{\rm shock}$. The thermal energy in the cold phase is
negligible in comparison, but in the hot phase it is substantial, $6
\times 10^{56}$ erg; however, on the shock model, about half of this
energy is not the result of AGN input, but is the internal energy of
the unshocked hot gas. Conservatively we take the magnetic field
  strength in the hot gas to be at the lower limit that we derived
  from the depolarization calculations, in which case its energy
  density makes only a small contribution to the total. Finally, the
minimum energy stored in the lobes is $\sim 3 \times 10^{56}$ erg.
Putting all this together (Table \ref{energies}), the AGN must have
supplied at least $10^{57}$ erg over its lifetime. The total estimated
work done by the jet in heating and accelerating the various phases of
the environment is comparable to (a factor $\sim 2$ larger than) the
{\it minimum} energy stored in the radio lobes, so that it is
plausible that there is rough equality between work done on the
environment and lobe energy -- this is in agreement with what is seen
in numerical simulations of FRIIs that drive shocks into their
environments (Hardcastle \& Krause, in prep.).

We can use the total energy calculated above to estimate the jet
power. The projected length of the NE jet is 2.2 kpc; thus for uniform
expansion\footnote{Radio sources probably do not expand
    uniformly, but for any realistic power-law dependence of lobe
    length $L$ on time $t$, the error in assuming $t = L/({\rm
      d}L/{\rm d}t)$ is of order unity; for example, in the
    self-similar models of Kaiser \& Alexander (1997), $L =
    Ct^{3/(5-\beta)}$, where $\beta$ is a power-law exponent of
    external density defined by Kaiser \& Alexander, with $1<\beta<3$
    from observation: then $L/({\rm d}L/{\rm d}t) = (5-\beta)t/3$, so
    the factor by which an assumption of uniform expansion is in error
    is between 2/3 and 4/3.} and given the observed radial velocity
  the lifetime $\tau = d/v = 8 \tan \theta$ Myr, or 3.5 Myr for
  $\theta = 24^\circ$ as estimated above, which gives a required
  time-averaged jet power of $10^{43}$ erg s$^{-1}$. Is this a
  reasonable kinetic luminosity for 3C\,305? H10 estimated a jet power
  around 30 times higher for 3C\,171 on a very similar basis, but
  3C\,171 is a much more powerful source in all other
  respects\footnote{Radio luminosity comparisons in this section use
    the 178-MHz flux densities and low-frequency spectral indices from
    Laing, Riley \& Longair (1983), adjusted to the flux scale of
    Roger, Bridle \& Costain (1973): see
    http://3crr.extragalactic.info/.} -- its low-frequency radio
  luminosity is about a factor 50 higher, for example -- so these
  numbers seem reasonable from that point of view. A more surprising
  comparison comes from considering FRI sources. For example, Croston
  \etal\ (2009) show that the jet power (including work done on the
  environment, but considering only one jet) of the closest FRI, Cen
  A, is $\sim 10^{43}$ erg s$^{-1}$. In a very different way, based on
  their detailed study of the dynamics of the jet combined with X-ray
  observations of its environment, Laing \& Bridle (2002) determine a
  jet power for the archetypal twin-jet FRI 3C\,31, a factor $\sim 6$
  less luminous at low frequencies than 3C\,305, of $10^{44}$ erg
  s$^{-1}$. In other words, there appear to be FRI jets whose kinetic
  powers are comparable to, or even substantially exceed, that of the
  FRII-like 3C\,305 (with its well-collimated, apparently relativistic
  jets and bright hotspots). Should this come as a surprise? Probably
  not, since we know that the FRI/FRII transition is both
  observationally and theoretically a strong function of environment
  (Ledlow \& Owen 1996; Bicknell 1995), and 3C\,31 at least inhabits a
  much richer environment than 3C\,305, one in which much of the
  kinetic power of the jet must be expended in accelerating (and
  possibly heating) entrained external material. However, the
  comparison of 3C\,31 and 3C\,305 clearly shows the danger of
  inferring jet powers directly from radio observations.

One remaining puzzle is why this luminosity is so much less than the
line luminosity estimated by H82, $10^{44}$ erg s$^{-1}$. H82
themselves noted that there are problems with such a high
emission-line luminosity if the optical line-emission regions are
ionized by the X-ray-emitting plasma, given that even their upper
limit on X-ray emission was $<10^{42}$ erg s$^{-1}$: this problem is
made worse by our detection of X-ray emission with a bolometric
luminosity of $4.4 \times 10^{41}$ erg s$^{-1}$, as we noted in
Section \ref{regions}. But our estimates of jet power here, coupled
with our belief that all the radiation seen from the system is a
result of an interaction between the jet and the external medium, add
a further difficulty: it is very hard to see how such a high line
luminosity can be powered by a jet that is on average carrying an
order of magnitude less power. In fact, it seems very likely that the
emission-line luminosity that H82 derive is a significant
overestimate. They derive an $E(B-V)$ of 0.85 from the large Balmer
decrement they observe, which is confirmed by later observations, e.g.
those of Buttiglione \etal\ (2009); but, as shown by Tadhunter
\etal\ (2005), the stellar continuum in 3C\,305 is dominated by an
intermediate-age stellar population with strong Balmer line
absorption, and, as this is much stronger for H$\beta$ than H$\alpha$,
large apparent H$\alpha$/H$\beta$ ratios are expected without some
correction for the underlying continuum, which was not carried out by
H82. Although we know from optical imaging observations (e.g. Martel
\etal\ 1999) that there is dust on scales comparable to the
emission-line regions, so that some reddening is expected, the
$E(B-V)$ values measured for the dusty narrow-line regions of other
radio galaxies tend to be $\sim 0.3$ -- 0.5 (e.g.\ Tadhunter
\etal\ 1994; Robinson \etal 2000), and it is unlikely that the value
for 3C\,305 greatly exceeds these. If we simply assume the Galactic
$E(B-V) \approx 0.03$ (from the dust maps of Schlegel \etal\ 1998)
then the line luminosity would come down to $\la 10^{43}$ erg
s$^{-1}$, and a more realistic correction for the reddening expected
from 3C\,305 would probably increase this by at most a factor of a
few. This luminosity is still high compared to the X-ray emission but
is no longer inconsistent with our estimates of jet and AGN power;
thus it seems plausible that some combination of shock-ionization from
the jet and photoionization from the AGN can give rise to the line
emission that we see.

\section{Summary and conclusions}
\label{conclusions}

We have presented new, sensitive X-ray and radio observations of the
nearby, peculiar radio galaxy 3C\,305. Our key results are as follows:
\begin{itemize}
\item We have argued based on the morphology of the observed X-rays,
  and the properties of the X-ray emission attributed to the hidden
  AGN, that the extended X-ray-emitting plasma is shock-ionized by the jet.
\item We have shown that we can construct a self-consistent model in
  which the X-ray-emitting plasma is, as found
  previously with 3C\,171, plausibly the material responsible for the
  observed depolarization in the radio, and used this to place a
  limit, $B>1.6$ nT, on the magnetic field in the external medium.
\item Combining the physical conditions estimated from the X-ray and
  depolarization analysis with what is known about the source dynamics
  from optical emission-line and neutral-hydrogen studies, we are able
  to make a complete (if model-dependent) estimate of the energy
  budget of the radio source. The X-ray-emitting phase dominates the
  energetics of the (known) phases of external gas, but warm and cold
  gas also make significant contributions. The work done on the
  various phases of the environment is comparable to (a factor $\sim
  2$ larger than) the minimum energy stored in the radio lobes.
\item From the work done on the environment and the minimum energy
  calculation we estimate a jet power for the source of
  $10^{43}$ erg s$^{-1}$, accounting for all known phases of the
  environment with which the jet is interacting.
\end{itemize}

Our results above demonstrate the value of a complete multi-wavelength
view of a radio source in understanding and quantitatively assessing
its environmental impact. The (model-dependent) energetic constraints
and kinetic luminosities we have derived require both the X-ray
observations (since the hot gas dominates the mass budget and has a
high energy density) and the optical/H{\sc i} measurements (since we
need velocities to estimate the kinetic energy). Detailed
emission-line or H{\sc i} studies alone, while crucial to tell us
about the kinematics, may be failing to see a
substantial fraction of the energy supplied by the AGN (e.g. Holt
\etal\ 2011). While there are
relatively few low-redshift targets for X-ray investigations of
radio-galaxy-driven outflows (Section
\ref{intro}, and below) we hope to apply these techniques to high-redshift
objects in the future. A direct test of the crucial assumption in our
  modelling above -- that the X-ray-emitting gas is outflowing at
  speeds comparable to the other phases -- will have to await
  high-sensitivity observations with calorimeters in the more distant future.

Finally, we return to the question of why 3C\,305, and its more
powerful counterpart 3C\,171, are so atypical for low-redshift objects
in their radio morphology and their relationship with the extended
emission-line and X-ray gas. In H10 we argued that the key feature
that distinguishes these objects from more typical radio galaxies of
the same jet/radio power is a strong interaction between the jet and
{\it cold} gas aligned along the radio axis, since we know that
considerably larger amounts of hot gas, even if asymmetrically
distributed, do not prevent the formation of a typical FRII radio
galaxy. This picture seems to be consistent with everything we see in
3C\,305. The smoking gun for this model in 3C\,305 is the presence of
neutral hydrogen in outflow along the jet axis (Morganti
\etal\ 2005a). This is very hard to understand in isolation -- how can
the jet drive neutral hydrogen without disrupting it? -- but easier
once we realise that the H{\sc i} is only a small fraction of the
outflowing gas. In this picture, then, the jet in 3C\,305 has driven a
shock into a medium which had originally had a much larger mass of
H{\sc i} and other cold/warm material, aligned along the jet axis and
embedded in the hot-gas environment expected for a massive elliptical
galaxy. The shock has heated much of the cold gas to high temperatures
and shredded the rest, and the remaining H{\sc i} and other warm-phase
material (e.g. the optical emission-line clouds) are being carried
along within the outflowing X-ray emission. As noted above, this may
help to explain the low abundance derived from APEC fitting to the hot
gas. The X-ray `wings', in this picture, must then be shocked material
which has been driven back towards the centre of the galaxy by the
expanding radio lobes. A somewhat similar scenario (with a similar
impact on radio morphology) is seen in the simulations of Wagner \&
Bicknell (2011). As we remarked in H10, it seems likely that the
rarity of similar objects at low redshift (in contrast to the much
more typical `classical double' morphology) reflects the rarity of
gas-rich mergers in the massive elliptical hosts of local radio
galaxies, but this type of interaction is almost certainly both more
common and energetically more significant in the earlier universe.

\section*{Acknowledgements}

We thank the referee, Clive Tadhunter, for constructive comments
  that have helped us to improve the paper. MJH thanks the Royal
Society for a Research Fellowship which supported the early parts of
this work. The work at SAO was supported by NASA grant G01-12133A. FM
acknowledges the Fondazione Angelo Della Riccia for the grant awarded
to support his research at SAO during 2011 and the Foundation
BLANCEFLOR Boncompagni-Ludovisi, n\'ee Bildt, for the grant awarded to
him in 2010. The National Radio Astronomy Observatory is a facility of
the National Science Foundation operated under cooperative agreement
by Associated Universities, Inc.

\appendix

\section{Error estimates for hardness ratios}

In the limit of large numbers of counts, error estimates for hardness
ratios can be derived using standard propagation of error techniques.
However, we are using hardness ratios precisely because we are too
photon-limited to do anything else (e.g. detailed spectral analysis),
particularly in the hard band, and so we have to consider
uncertainties more carefully. Our analysis differs from that of
Gehrels (1986) in being explicitly Bayesian; we focus not on the
`errors' on the measured values but on the uncertainties on the
inferred quantities of interest. Park \etal\ (2006) have investigated
the Bayesian approach to this problem in some detail, but the
generality of their approach leads to a quite complicated solution.
Here we describe a simple analysis which is valid in the limit in
which the background can be neglected, and which gives identical
results to those of Park \etal\ in this limit\footnote{We tested this
  with the code made available at
  http://hea-www.harvard.edu/AstroStat/BEHR/ .}.

Poisson statistics are described by the familiar equation
\begin{equation}
p(n|\mu) = \frac{\mu^n e^{-\mu}}{n!}
\end{equation}
where $\mu$ is a known expected number of counts and $n$, an integer,
is the number of counts actually observed. We are in the position,
though, of observing a given number of counts $n$ and wanting to infer
the underlying expected number $\mu$. In this case, by Bayes' theorem
(and assuming an uninformative, though improper, prior) we know that
\begin{equation}
p(\mu|n) = \mu^n e^{-\mu}/\int_0^\infty \mu^n e^{-\mu} {\rm d}\mu
\end{equation}
The normalizing integral here is just the gamma function,
$\Gamma(n+1)$, so we can write down the posterior probability
distribution of the expected number of counts $\mu$:
\begin{equation}
p(\mu|n) = \frac{\mu^n e^{-\mu}}{\Gamma(n+1)}
\label{pd}
\end{equation}
This probability distribution is known as the gamma distribution. The
maximum-likelihood value of $\mu$ (the posterior mode) is, as
expected, $n$. The mean of the distribution is $n+1$ and variance
$n+1$, and it converges to a Gaussian in the limit of large $n$, so we
then recover the standard result that the best estimate of $\mu$ is
$\mu \approx n \pm \sqrt{n}$. In general, though, we would want to
estimate the errors on $\mu$ by defining a `credible interval'
analogous to those used in the case of a Gaussian: for example, the
$1\sigma$ error range $\mu_l$ -- $\mu_h$ could be defined by
\begin{equation}
\int_{\mu_l}^{\mu_h} p(\mu|n) {\rm d}\mu = \erf(1/\sqrt{2}) = 0.6823
\label{errors}
\end{equation}
subject to the constraint that $\mu_h-\mu_l$ takes its minimal value.

Now consider a hardness ratio $H$, defined as in the text. In an
experiment in which two sets of numbers of counts $\mu_1$ and $\mu_2$ could be
measured with arbitrary precision, we would expect
\begin{equation}
H = 1 + \frac{\mu_2 - \mu_1}{\mu_2 + \mu_1}
\label{hr-mu}
\end{equation}
However, in fact, our measurements of numbers of counts $n_1$ and
$n_2$ give us probability distributions for $\mu_1$ and $\mu_2$ 
according to eq.\ \ref{pd} above. Consequently there is a probability
distribution for $H$, $p(H|n_1,n_2)$. Our best estimate of $H$
would, again, be $\int H p(H|n_1,n_2){\rm d}H$ and the error
estimates should be derived in a way analogous to that given in
eq.\ \ref{errors}.

We cannot easily write down the probability distribution for $H$
analytically, but it is easy to determine it by Monte Carlo
simulation. We know $p(\mu_1|n_1)$ and $p(\mu_2|n_2)$ from
eq.\ \ref{pd} and so, drawing values of $\mu_1$ and $\mu_2$ from this
distribution (we use the GNU Scientific Library function
gsl\_ran\_gamma), we can calculate $H$ for each pair from
eq.\ \ref{hr-mu}. This gives us a set of $H$ values drawn from $p(H)$,
from which we can trivially calculate the Bayesian estimator of $H$
(the posterior mean) and the $1\sigma$ credible interval. Credible
interval calculations for a confidence limit $CL$ will be accurate so
long as $N(1-CL) \gg 1$: we use $CL=0.6823$ and $N = 50000$ in the
results presented in the text.

We should note that the use of the posterior mean
can give some slightly non-intuitive results in the limit of small
$n_1,n_2$. For example, consider the case where $n_2 = 0$, $n_1 > 0$.
Simply calculating the $H$ values from the counts, we expect $H = 0$.
But for $n_2=0$, the posterior probability distribution of $\mu_2$,
eq. \ref{pd}, reduces to the exponential distribution, whose mean is
not zero: in other words, given our assumptions, the Bayesian
estimator of $\mu_2 >0$ even though we measure no counts and though
the maximum-likelihood estimator of $\mu_2 = 0$. Consequently we
expect the Bayesian estimator of $H$, and its $1\sigma$ confidence
limits, to be $>0$. As it is computationally simpler to calculate the
Bayesian estimator and its credible interval from the Monte Carlo than
it is to do the same for the maximum-likelihood estimate, we plot the
Bayesian estimator of $H$ in Fig.\ \ref{hr-results} in the text.

\end{document}